\documentclass[12pt]{JHEP3}

\usepackage{amssymb,epsfig}

\newcommand{\beq}{\begin{equation}}
\newcommand{\eeq}{\end{equation}}

\newcommand{\Omegam}{\Omega_{\mathrm{M}}}
\newcommand{\Omegab}{\Omega_{\mathrm{B}}}

\newcommand{\Omegade}{\Omega_{\mathrm{DE}}}

\newcommand{\omegam}{\omega_{\mathrm{M}}}
\newcommand{\omegab}{\omega_{\mathrm{B}}}
\newcommand{\nsc}{n_{\mathrm{s}}}
\newcommand{\dr}{\Delta_{\mathcal{R}}^{2}}
\newcommand{\wzero}{w_{0}}
\newcommand{\wa}{w_{\mathrm{a}}}

\newcommand{\wgrow}{w_{\mathrm{g}}}
\newcommand{\wdist}{w_{\mathrm{d}}}
\newcommand{\wdiff}{w_{\mathrm{diff}}}

\newcommand{\Rvir}{R_{200\mathrm{m}}}

\newcommand{\Pktom}[2]{P_{\kappa}^{\mathrm{#1}\mathrm{#2}}}
\newcommand{\Pkobs}[2]{\bar{P}_{\kappa}^{\mathrm{#1}\mathrm{#2}}}

\newcommand{\ellmax}{\ell_{\mathrm{max}}}
\newcommand{\wlw}[1]{W_{\mathrm{#1}}}
\newcommand{\Da}{D_{\mathrm{A}}}

\newcommand{\ntomo}{N_{\mathrm{TOM}}}

\newcommand{\zp}{z_{\mathrm{p}}}

\newcommand{\fsky}{f_{\mathrm{sky}}}
\newcommand{\gi}{\langle \gamma^2 \rangle}

\newcommand{\dd}{\mathrm{d}}

\title{The Influence of Galaxy Formation Physics on Weak Lensing 
Tests of General Relativity}

\author{Andrew P. Hearin and Andrew R. Zentner\\
Department of Physics and Astronomy, University of Pittsburgh,\\
Pittsburgh, PA 15260 USA\\
E-mail: \email{aph15+@pitt.edu,zentner+@pitt.edu}}

\abstract{
Forthcoming projects such as the Dark Energy Survey, 
Joint Dark Energy Mission, and the Large Synoptic Survey 
Telescope, aim to measure weak lensing shear correlations with 
unprecedented accuracy.  Weak lensing observables are sensitive to 
both the distance-redshift relation and the growth of structure 
in the Universe.  If the cause of accelerated cosmic expansion is dark energy 
within general relativity, both cosmic distances and structure growth are 
governed by the properties of dark energy.  Consequently, one may use lensing to check for 
this consistency and test general relativity.  After reviewing the 
phenomenology of such tests, we address a major challenge to such a program.  
The evolution of the baryonic component of the Universe is highly uncertain 
and can influence lensing observables, manifesting as 
modified structure growth for a fixed cosmic distance scale.  
Using two proposed methods, we show that one could be led 
to reject the null hypothesis of general relativity when it 
is the true theory if this uncertainty in baryonic processes is neglected.  
Recent simulations suggest that we can correct for baryonic effects using a 
parameterized model in which the halo mass-concentration relation is modified.   
The correction suffices to render biases small compared to statistical uncertainties.  
We study the ability of future weak lensing surveys to constrain 
the internal structures of halos and test the null hypothesis of 
general relativity simultaneously.  
Compared to alternative methods which null information from small-scales 
to mitigate sensitivity to baryonic physics, this internal calibration 
program should provide limits on deviations from general relativity that 
are several times more constraining.  Specifically, we find that limits on 
general relativity in the case of internal calibration are degraded by only 
$\sim 30\%$ or less compared to the case of perfect knowledge of 
nonlinear structure.  
}


\begin{document}


\section{Introduction}
\label{section:intro}
Current data indicate that the cosmological expansion is accelerating 
(e.g., Refs.~\cite{riess_etal98,perlmutter_etal99,tegmark_etal04,riess_etal04,
eisenstein_etal05,spergel_etal07,tegmark_etal06,astier_etal06,wood-vasey_etal07}).  
This accelerating expansion is one of the most profound discoveries of the past 
decade and points to a fundamental gap in our understanding of the universe.  
The cause of the accelerating 
expansion has been dubbed dark energy.  Indeed dark energy is simply a name for the 
fundamental puzzle that is the nature of cosmic acceleration and an enormous amount of effort is 
now being devoted to shedding light on the causative agent of cosmic acceleration.  In broadest terms, two 
options have been put forth to describe accelerated cosmic expansion.  The first is that 
cosmic acceleration is caused by some as yet unidentified contribution to the stress 
energy of the universe.  This option includes vacuum energy (observationally indistinguishable 
from a cosmological constant) and dynamical models of dark energy.  The second option is that 
gravity deviates from the general relativistic description on large scales, an option we 
refer to as modified gravity.

Viable options to general relativity for which definite predictions 
have been made are few and far between, so many authors have suggested that a fruitful way to 
apply forthcoming data will be to check for the mutual consistency of different observable 
phenomena with the predictions of general relativity 
\cite{zhang_etal05,linder05,zhan_knox06,wang_etal07,huterer_linder07,linder_cahn08,zhan_etal08,mortonson_etal08}.  
The basic idea is that one may obtain observational handles on both the distance-redshift 
relation and the growth rate of cosmic expansion.  Within general relativity, both of these 
can be predicted from the same dark energy parameters, so it is possible to measure, for example, 
distance and then check for consistency with the growth of cosmic structure.  Forthcoming 
weak gravitational lensing surveys will provide the most powerful means to probe the matter distribution 
in the universe directly and are an indispensable piece of any such consistency 
check \cite{zhang_etal05,huterer_linder07}.  A significant literature already exists detailing the 
power of tomographic weak lensing to constrain dark energy parameters, including 
Refs.~\cite{hu_tegmark99,hu99,huterer02,heavens03,refregier03,refregier_etal04,song_knox04,takada_jain04,
takada_white04,bernstein_jain04,dodelson_zhang05,albrecht_etal06,zhan06,zentner_etal08}.  In fact, 
the effectiveness of any given matter fluctuation to serve as a lens is sensitive to the distance 
scale of the universe and so weak gravitational lensing alone will provide a very powerful 
check for the consistency of general relativity.  In this manuscript, we study the effectiveness 
of such a consistency check in light of recent studies that indicate that predictions for weak 
lensing observables will contain significant inherent uncertainties due to the poorly-understood 
behavior of the baryonic component of the universe \cite{white04,zhan_knox04,jing_etal06,rudd_etal08,zentner_etal08}.

The strategy for checking the consistency of general relativity and searching for a sign of 
modified gravity outlined in the previous paragraph seems simple enough.  However, making 
predictions for weak lensing power spectra is fraught with numerous practical difficulties.  
Weak lensing will use information on scales where density fluctuations are well beyond the 
linear regime (relative overdensities $\delta \gtrsim 1$).  To utilize this information 
at the level that forthcoming observational programs will permit, 
numerical simulations must be able to predict the nonlinear matter power spectra to better 
than a percent \cite{huterer_takada05,huterer_etal06}.  Sufficient precision should be 
achievable by brute force using dissipationless, $N$-body 
numerical simulations that treat only the evolution of the 
cosmic density field under gravity only 
(e.g., Refs.~\cite{heitmann_etal05,heitmann_etal08,heitmann_etal08b,heitmann_etal09}).  
The most challenging obstacle to such 
precise predictions is the significant influence that baryonic processes have on 
matter power spectra on scales of interest.  This has been pointed out in both 
analytic \cite{white04,zhan_knox04} and numerical studies \cite{jing_etal06,rudd_etal08}.

The problem posed by baryonic processes is severe because they cannot be 
modeled directly and all such calculations rely on relatively poorly-constrained, 
effective models that approximate the net, large-scale, effects of processes that 
occur on scales far below the numerical resolution that may be achieved with any simulation.  
Ref.~\cite{rudd_etal08} shows that the influences of baryons may be large compared with the 
statistical uncertainties expected of future surveys and that different treatments of baryonic 
physics lead to notably different matter power spectra on relevant scales.  
Ref.~\cite{zentner_etal08} studied a parameterized model 
for baryonic influences that could treat all of the simulations of 
Ref.~\cite{rudd_etal08} and showed that such a parameterized model could be calibrated self-consistently 
using weak lensing data alone, yielding meaningful constraints on both the dark energy and the 
effective models of baryonic physics.  However, extending the 
results of Ref.~\cite{zentner_etal08} to consistency checks of general relativity requires care.    
Baryonic processes manifest as an inability to predict the evolution of the density 
field even for a fixed cosmic expansion history.  This is, in part, what Ref.~\cite{zentner_etal08} 
relied upon in their self-calibration program and this is precisely what proposed tests of 
the consistency of general relativity rely upon.

The strategy behind all efforts to study dark energy using gravitational lensing 
relies on the ability to produce reliable $N$-body simulations.  In fact, most studies 
implicitly assume such a simulation campaign will be performed prior to any data analysis 
and proceed to estimate the power of forthcoming experiments using contemporary fitting 
formulae.  In fact these fitting formulae are not sufficiently precise to treat 
forthcoming data, they may be subject to fundamental limitations 
\cite{lukic_etal07,tinker_etal08,robertson_etal08,heitmann_etal08b,heitmann_etal09}, 
and, aside from a handful of specific cases, they have not been generalized to treat dark energy.  
A numerical campaign will be necessary to address dark energy and addressing observables using 
dissipationless $N$-body simulations is a challenging, but tractable problem 
\cite{heitmann_etal08,heitmann_etal09}.  
The results of $N$-body simulation campaigns represent precise solutions to the idealized problem of 
computing the density or lensing fields in a variety of cosmologies absent baryonic physics.  
As with dark energy constraints, the strategy of self-calibrating the net influences of baryons is 
based on the assumption that such a set of simulations without baryons are available and that a simple set of 
prescribed corrections can be applied to the $N$-body results to describe baryonic 
effects.  The reason is that direct numerical calculations that treat the physics of 
baryons are not achievable in the foreseeable future, given current computational limitations.
This is the context of the present paper.

In this paper, we extend the results of Ref.~\cite{zentner_etal08} 
to address consistency checks of general relativity.  
We present a brief pedagogical discussion of the manner in which 
independent modifications to the distance scale of the universe and 
the growth rate of structure produce non-degenerate changes in weak lensing 
statistics, enabling weak lensing alone to serve as a powerful consistency 
check for general relativity.  We then move on to discuss the interplay between 
modifications to convergence spectra caused by baryonic processes and the 
dark energy.  We show explicitly that our current ignorance regarding the 
evolution of the baryonic component of the universe, its influence on the 
evolution of inhomogeneities of dark matter, and the process of galaxy formation 
places a severe limitation on our ability to test the consistency of general relativity.  
To be specific, we consider two tests.  In one test, the growth of structure and the 
cosmic distance scale are assumed to arise from two different effective equations of state 
$\wgrow$ and $\wdist$, and we test our ability to rule out the hypothesis that the two are 
equal as they would be in general relativity.  
If unaccounted for, our limited ability to predict convergence spectra can lead to 
biases that drive these two parameters to disagree at levels as large as $\sim 8\sigma$ 
(depending upon details) when general relativity is the correct description of gravity.  
We perform a similar test 
on the gravitational growth index parameter $\gamma$, introduced in Ref.~\cite{linder05}, 
and likewise find significant biases if baryonic effects are not treated.  
We show that it is possible to eliminate these biases by disregarding the small-scale 
shear information from forthcoming surveys, but the cost is a 
factor of $\sim 2-4$ degradation in the constraining 
power of such surveys.  Lastly, we study the ability of forthcoming surveys to 
test general relativity and constrain baryonic processes simultaneously within 
a single data set.  We show that this is a promising option as the biases can be eliminated 
at a cost of only $\sim 20\%$ in parameter constraints.  This implies that significant 
effort should be devoted to developing robust parameterizations of the nonlinear evolution 
of structure as we begin to realize the next generation of imaging surveys.  These methods should 
extend current techniques by including realistic descriptions of the 
effects of baryons that are not so general as to be completely arbitrary but 
do allow sufficient freedom so as to reflect our ignorance of the influence of 
the baryonic sector.

In the following section, we describe our methods, including our parameterizations of 
cosmological expansion and structure growth.  In \S~\ref{section:results}, we present 
the results of our study.  We begin in \S~\ref{sub:consistency} by illustrating the 
power of these consistency checks.  This section contains a compilation of facts and 
figures found dispersed throughout the existing literature.  
In \S~\ref{sub:biases}, we show how neglect of 
baryonic processes may lead to biases that would mimic an inconsistency in 
general relativity in simpler treatments of weak lensing observables.  In \S~\ref{sub:calibration}, 
we present our results that show that additional parameter freedom may be added to 
predictions of the growth of structure to account for the influence of baryons and make 
such consistency checks robust to nonlinear processes.  In \S~\ref{section:discussion}, 
we discuss our results and we summarize our efforts in \S~\ref{section:conclusions}.

\section{Methods}
\label{section:methods}

\subsection{Weak Lensing Observables}
\label{subsection:observables}

We treat weak lensing observations as a consistency check of general relativity.  Aside 
from our parameterization of dark energy, we perform all calculations as described in 
detail in Ref.~\cite{zentner_etal08}.  We summarize those calculations here and 
refer the reader to this reference for more detail.  We consider number density-weighted 
convergence power spectra and cross-spectra from 
$\ntomo=5$ tomographic redshift bins as our primary observables:
%
\beq
\label{eq:pkij}
\Pktom{i}{j}(\ell) = \int \dd z  \frac{\wlw{i}(z)\wlw{j}(z)}{H(z)\Da^2(z)}P(k=\ell/\Da,z).
\eeq
The five photometric redshift bins are spaced equally in redshift between a 
minimum photometric redshift of $\zp^{\mathrm{min}}=0$ and a maximum $\zp^{\mathrm{max}}=3$.  
Previous work showed that parameter constraints are saturated with 
$\ntomo=5$ and that further binning is unnecessary \cite{ma_etal06,zentner_etal08}.  
We have verified that this remains so in the models we consider.  
In Eq.~(\ref{eq:pkij}), $\Da$ is the angular diameter distance, 
$H(z)$ is the Hubble expansion rate, $P(k,z)$ is the matter power spectrum 
at wavenumber $k$ and redshift $z$, and $\wlw{i}(z)$ are the lensing weight 
functions.  Given the true redshift distribution of sources in bin $i$, 
$\dd n_i/\dd z$, the lensing weight is 
\beq
\label{eq:weight}
\wlw{i}= \frac{3\Omegam H_0^2}{2} (1+z)\Da \int \frac{\Da(z,z')}{\Da(z')} \ \frac{\dd n_i}{\dd z'} \ \dd z'.
\eeq

We assume that the true redshift distribution of sources is 
\beq
\label{eq:sources}
\dd n(z)/\dd z = 4\bar{n}(z/z_0)^2\exp[-(z/z_0)^2]/\sqrt{2\pi z_0^2},
\eeq
where $\bar{n}$ represents the total density of source galaxies per unit 
solid angle, which varies from survey to survey, and we assume 
$z_0 \simeq 0.92$ so that each survey has a median redshift of $z_{\mathrm{med}}=1$.  
We assume that the probability of a photometric redshift $\zp$ given a 
spectroscopic redshift $z$ is Gaussian with a mean value of $\zp$ given by $z$ (no bias) 
and a dispersion $\sigma_z=0.05(1+z)$, and compute $\dd n_i/\dd z$ for each photometric 
redshift bin as described in Ref.~\cite{ma_etal06}.

We estimate the constraining power of forthcoming weak lensing surveys using 
the Fisher information matrix.  Indexing the observables of Eq.~(\ref{eq:pkij}) 
by a single label, we write $\mathcal{O}_{\mathrm{A}}=\Pktom{i}{j}$, where 
each $i,j$ map onto a unique $A$, and the Fisher matrix of weak lensing 
observables is 
%
\beq
\label{eq:fisher}
F_{\alpha \beta}=\sum_{\ell_{\mathrm{min}}}^{\ellmax} (2\ell+1)\fsky 
\sum_{\mathrm{A,B}} \frac{\partial \mathcal{O}_{\mathrm{A}}}{\partial p_{\alpha}} 
[C^{-1}]_{\mathrm{AB}} 
\frac{\partial \mathcal{O}_{\mathrm{B}}}
{\partial p_{\beta}} + F_{\alpha \beta}^{\mathrm{P}}. 
\eeq
The $p_{\alpha}$ represent the parameters of the model, Greek indices label model parameters, 
capital Latin indices label unique observables, and lower-case Latin indices label tomographic 
redshift bins.  $\fsky$ is the fraction of sky covered by an experiment, $C_{\mathrm{AB}}$ is 
the covariance matrix of observables, and the sum begins at $\ell_{\mathrm{min}}=2/\sqrt{\fsky}$ 
and runs to some $\ellmax$.  We generally take $\ellmax=3000$ to ensure that we only 
use scales where assumptions of weak lensing and Gaussian statistics are 
valid \cite{white_hu00,cooray_hu01,vale_white03,dodelson_etal06,semboloni_etal06}.  
However, we present many of our primary results (Figures~\ref{fig:consist},
~\ref{fig:wbiases},~\ref{fig:gbiases},~\ref{fig:cal}) as a function of $\ellmax$ 
so results for alternative choices are simple to extract.  
Observed spectra contain both signal and noise, 
$\Pkobs{i}{j} = \Pktom{i}{j} + n_{\mathrm{i}} \delta_{\mathrm{ij}} \gi$, where 
$n_{\mathrm{i}}$ is the surface density of source galaxies in bin $\mathrm{i}$ and $\gi$ is the 
intrinsic source galaxy shape noise.  We follow convention in setting $\gi=0.2$ and 
allowing differences in shape noise between different observations to be absorbed into $n_{\mathrm{i}}$.  
The covariance between observables $\Pkobs{i}{j}$ and $\Pkobs{k}{l}$ is 
$C_{\mathrm{AB}}=\Pkobs{i}{k} \Pkobs{j}{l} + \Pkobs{i}{l} \Pkobs{j}{k}$, 
where $\mathrm{i}$ and $\mathrm{j}$ map to $\mathrm{A}$, 
and $\mathrm{k}$ and $\mathrm{l}$ map to $\mathrm{B}$.

The inverse of the Fisher matrix is an estimate of the parameter covariance near 
the maximum of the likelihood.  The measurement error on parameter $\alpha$ marginalized 
over all other parameters is
%
\beq
\label{eq:ferror}
\sigma(p_{\alpha})=[F^{-1}]_{\mathrm{\alpha \alpha}}.
\eeq
The second term in Eq.~(\ref{eq:fisher}) incorporates Gaussian priors on model parameters.  We assume 
modest priors on several cosmological parameters individually, so that 
$F^{\mathrm{P}}_{\alpha \beta} = \delta_{\alpha \beta}/\sigma_{\alpha}^2$, where 
$\sigma_{\alpha}$ is the assumed prior uncertainty on parameter $\alpha$.  
Unless otherwise stated, when we refer to the uncertainty in a parameter or subset of 
parameters, we are referring to the uncertainty in the parameters under discussion after 
marginalizing over the remaining parameters of the model.  
This formalism also provides an approximation for biases in cosmological 
parameter estimators due to unknown, untreated, systematic offsets in 
observables.  Let $\Delta \mathcal{O}_{\mathrm{A}}$ be the 
difference between the true observable and the prediction for that observable 
absent the systematic.  The induced bias in the estimator of 
parameter $\alpha$ due to the neglect of the systematic offset is 
%
\beq
\label{eq:bias}
\delta p_{\alpha} = \sum_{\beta} [F^{-1}]_{\alpha \beta} 
\sum_{\ell} (2\ell+1) \fsky \sum_{\mathrm{A,B}} 
\Delta \mathcal{O}_{\mathrm{A}} [C^{-1}]_{\mathrm{AB}} 
\frac{\partial \mathcal{O}_{\mathrm{B}}}{\partial p_{\beta}}.
\eeq

We model the matter power spectrum using the phenomenological halo model 
\cite{scherrer_bertschinger91,ma_fry00,seljak00} (for a review see Ref.~\cite{cooray_sheth02}).  
We use the particular implementation of Ref.~\cite{zentner_etal08}.  This model utilizes 
standard fitting formulae for halo abundance and halo bias \cite{sheth_tormen99}.  It is 
known that existing fitting formulae are not yet sufficiently precise to treat forthcoming 
data sets \cite{lukic_etal07,tinker_etal08}, but it is likely that this can be overcome.  
Our approach is premised on the idea that dissipationless simulation programs will be carried 
out to calibrate these quantities to the necessary precision (or some equivalent 
strategy that utilized the simulation data directly), and that baryonic processes 
are the only processes that are not treated with sufficient precision.  
The non-standard modification to the halo model that we consider concerns the 
distribution of matter within dark matter halos.  
Typically, it is assumed that on average the mass density within a halo is 
described by the standard Navarro, Frenk, \& White (NFW) density profile \cite{navarro_etal97}, 
$\rho \propto (cr/\Rvir)^{-1}(1+cr/\Rvir)^{-2}$, where $\Rvir$ is the halo 
virial radius which we define to contain a mean density of $200$ times the 
mean density of the universe, the normalization is set by the fact that the 
mass profile must integrate to contain the total virial mass $m$ within $\Rvir$, 
and $c$ is the halo concentration which sets 
the radius of the transition between the two power laws as $r_{\mathrm{s}} \sim \Rvir/c$.

The standard practice is to set the average concentration of a halo of mass 
$m$ according to a phenomenological law derived from dissipationless $N$-body 
simulations, such as 
\beq
\label{eq:cofm}
c(m,z)=c_0 [m/m_{\star,0}]^{-\alpha}(1+z)^{-\beta},
\eeq
where $c_0 \approx 10$, $\alpha \approx 0.1$, $\beta \approx 1$ \cite{bullock_etal01,maccio_etal07,neto_etal07}, 
and $m_{\star,0}$ is the mass of a typical object collapsing at $z=0$.  We neglect the 
spread in halo concentrations at fixed mass because this is gives rise to a negligible 
effect on the scales we consider, and on smaller scales this dispersion is degenerate 
with an overall shift in the concentration-mass relation \cite{cooray_hu01,dolney_etal04}.  
The shortcoming of Eq.~(\ref{eq:cofm}) is that it describes halos in dissipationless $N$-body simulations that neglect 
the physics of baryons.  Fortunately, Ref.~\cite{zentner_etal08} demonstrated that 
adopting a modified concentration-mass law within the halo model suffices to model 
the convergence power spectra predicted by the baryonic simulations of Ref.~\cite{rudd_etal08} 
to a level where biases in inferred dark energy parameters are less than 10\% of their statistical 
uncertainties for forthcoming surveys.  We use this as a starting point for our preliminary 
estimate of the influence of baryonic processes on tests of the consistency of general 
relativity through weak gravitational lensing.

\subsection{Parameterized Tests of the Consistency of General Relativity}
\label{subsection:gravity}

As we mentioned in the introductory section, the ideal scenario would be to test parameterized 
families of theories for modified gravity that make specific and unique predictions; however, this 
is difficult at present.  Few if any viable alternatives to general relativity that contain a contemporary 
epoch of accelerated expansion have been identified.  Moreover, weak lensing requires some treatment 
of the evolution of density perturbations beyond the linear order of perturbation theory, and the 
nonlinear evolution of perturbations in models of modified gravity are not completely specified, 
nor have they been studied thoroughly.  We consider two different parameterizations that have been proposed to test for the 
consistency of general relativity within forthcoming data.  These do not represent complete, 
entirely self-consistent 
descriptions of any particular phenomenological alternative general relativity.  
We consider them as a pragmatic step toward testing the null hypothesis of gravity 
described by general relativity in the case of modest departures from the standard gravity.  
In both cases, we explore deviations to the linear growth of 
perturbations but assume that the relation between linear and nonlinear 
perturbations is unchanged.  In practice, this means that we assume that halos of dark matter 
form with the abundances and other gross properties that they otherwise would have in a standard-gravity 
treatment and continue to employ the halo model to predict convergence spectra [Eq.~(\ref{eq:pkij})] 
on nonlinear scales ($\ell \gtrsim 300$).  Modified gravity models require some environmental dependence 
to the gravitational force law so that gravity may deviate from general relativity on large scales (low density) 
yet satisfy local constraints on deviations from general relativity 
(in the high density environment of our galaxy) 
\cite{khoury_weltman04a,khoury_weltman04b,navarro_acoleyen06,navarro_acoleyen07,faulkner_etal07,hu_sawicki07a,oyaizu_etal08}.  
We do not consider any such modifications because there is no comprehensive treatment 
of the nonlinear evolution of perturbations in such theories and no phenomenological model 
akin to the halo model in the case of general relativity, though preliminary 
work in this direction has begun \cite{oyaizu_etal08}.

Our first test is to split 
the dark energy equation of state parameter into two distinct parameters.  The first parameter, 
$\wgrow$, is used in all calculations of the growth of density perturbations while the 
second, $\wdist$, is used in all calculations of the relationship between redshift and distance.  
The deflection of light is given by a sum of the Newtonian and curvature potentials and 
in the context of general relativity with zero anisotropic stress perturbation, these 
potentials are equal\footnote{We are admittedly a bit cavalier here as there are several 
different notational conventions in use, 
see Refs.~\cite{bardeen_etal83,lyth85,mukhanov_etal92,ma_bertschinger95,
hu_sawicki07a,hu_sawicki07b}; however, our contribution is not to explore the detailed 
behavior of the metric potentials in any specific theory (largely for lack of 
such options), so we forego a detailed discussion.  
In fact, we have already assumed the equality of the Newtonian and curvature potentials in 
order to derive Eq.~(\ref{eq:pkij}) for the convergence power 
spectrum in terms of the power spectrum of density fluctuations and 
which would otherwise be written in terms of the power spectra of the 
metric potentials directly.}.~
This equality means that perturbation growth and the distance-redshift relation are both determined by the 
evolution of the cosmic expansion rate $H(z)$, so that $\wgrow$ and $\wdist$ should be equal 
in general relativity and that an indication to the contrary may be a sign of modified gravity.  
Of course, any observational or theoretical systematic errors that drive the inferred values of 
$\wgrow \ne \wdist$ must be controlled and accounted for in order for this program to work.  
An alternative way to state our present aim is that we seek 
to assess the importance of the inability to predict the influence of the baryonic component 
of the universe on lensing power spectra for such consistency checks and explore a method 
to ensure the robustness of such a consistency check.  In this first test, we consider our 
parameter set for dark energy to consist of 
the present dark energy density in units of the critical density $\Omegade$, 
as well as the two equation of state parameters $\wdist$ and $\wgrow$.  
We take no priors on these parameters and set their 
fiducial values to $\Omegade=0.76$, $\wdist=\wgrow=-1$.

In the second case, we explore the gravitational growth index parameter $\gamma$, introduced by 
Linder in Ref.~\cite{linder05} and explored in Refs.~\cite{huterer_linder07,linder_cahn08}.  
Linder showed that a natural separation between the expansion history of the universe and 
the growth of perturbations could be achieved by taking the evolution of an overdensity 
$\delta$ to be given by 
\beq
\label{eq:delta_gamma}
\frac{\dd \ln \delta}{\dd \ln a} = \Omegam(a)^{\gamma}.  
\eeq
Our notation is such that $a=1/(1+z)$ is the cosmic scale factor, $\Omegam$ with no 
argument of scale factor or redshift is the current density of non-relativistic matter 
in units of the critical density, $\Omegam(a)=\Omegam a^{-3} H_0^2/H(a)^2$ with an explicit 
argument represents the evolution of the ratio of the matter density to the critical density, 
and $H_0 \equiv H(z=0) = 100 h$~km/s/Mpc is the present Hubble expansion rate.  
Eq.~(\ref{eq:delta_gamma}) holds for perturbations independent of scale or cosmic expansion 
history (aside from the dilution of $\Omegam(a)$.  In an enormous variety of dark energy 
models embedded within general relativity the growth index obeys 
\cite{linder05,huterer_linder07,linder_cahn08}
\begin{eqnarray}
\label{eq:gamma}
\gamma & = & 0.55 + 0.02[1+w(z=1)],\quad \quad \mathrm{for}\quad w<-1 \\
       & = & 0.55 + 0.05[1+w(z=1)],\quad \quad \mathrm{for}\quad w>-1, \nonumber
\end{eqnarray}
where $w(z)$ is the dark energy equation of state evaluated at redshift $z$.  Deviations 
from this relation therefore, may indicate non-standard gravity.  This treatment is 
accurate to better than a percent in specific cases such as the self-accelerating braneworld gravity 
model of Ref.~\cite{dvali_etal00} (so-called DGP gravity, which is now ruled out 
as an explanation of the cosmic acceleration, see Ref.~\cite{fang_etal08}, but still 
serves as a useful example of the utility of this parameterization), 
which yields $\gamma \simeq 0.69$, as well as scalar-tensor and $f(R)$ theories of gravity 
\cite{linder05,linder_cahn08}.  In this case, we consider the expansion history to be dictated 
by dark energy with abundance $\Omegade$ and a time-dependent effective equation of state 
$w(a)=\wzero+\wa(1-a)$ \cite{chevallier_polarski01}, 
while perturbation growth is dictated by $\gamma$.  We take no prior 
constraints on these parameters and perturb about the fiducial values 
$\Omegade=0.76$ (as in the first case), $\wzero=-1$, $\wa=0$, and $\gamma=0.55$.

\subsection{Cosmological Parameters and Future Lensing Surveys}
\label{subsection:experiments}

In the previous section, we described the parameters we use to describe deviations from 
general relativistic gravity.  To summarize, we take two test of the consistency of general 
relativity.  In both cases, we assume dark energy with a present density of $\Omegade=0.76$ in 
our fiducial model.  In the first case, we split the dark energy equation of state parameter into 
two pieces, one governing the growth of density inhomogeneities $\wgrow$, and one governing the 
distance redshift relation $\wdist$.  We take as a fiducial model $\wgrow=\wdist=-1$ and assume no 
priors on either parameter.  In the second case, we assume dark energy with an effective equation 
of state $w(a)=\wzero+\wa(1-a)$, with fiducial values $\wzero=-1$ and $\wa=0$.  In this case we 
assume inhomogeneity growth parameterized by $\gamma$ according to Eq.~(\ref{eq:delta_gamma}), 
and take the fiducial value of $\gamma=0.55$ as in general relativistic gravity with a cosmological 
constant causing contemporary acceleration.

Beyond the parameters that describe the dark energy/modified gravity sector, we 
consider four other cosmological parameters that influence lensing power spectra and may 
be degenerate with dark energy parameters.  These parameters and their fiducial values are:  
the non-relativistic matter density $\omegam \equiv \Omegam h^2 = 0.13$, the baryon density 
$\omegab = \Omegab h^2 = 0.0223$, the amplitude of primordial curvature fluctuations 
$\dr = 2.1 \times 10^{-9}$ (in practice we actually vary $\ln \dr$) at the 
pivot scale of $k_p=0.05$~Mpc$^{-1}$, and the power-law index of the spectrum of 
primordial density perturbations $\nsc = 0.96$.  We adopt conservative priors on each of these 
additional parameters that are comparable to contemporary constraints on each of these 
parameters individually \cite{komatsu_etal08}.  To be specific, we take prior constraints 
of $\sigma(\omegam)=0.007$, $\sigma(\omegab)=10^{-3}$, $\sigma(\ln \dr)=0.1$, and $\sigma(\nsc)=0.04$.  
Lastly, we allow three parameters that describe the effective concentrations of halos to 
be determined internally from the same data.  These are the parameters $c_0=13$, $\alpha=0.05$, and 
$\beta=1$ of the power-law halo mass-concentration relation in Eq.~(\ref{eq:cofm}).  The fiducial 
values we choose, $c_0 \approx 10$ and $\alpha \approx 0.1$, differ from those found in 
N-body simulations of structure formation [discussion after Eq.~(\ref{eq:cofm})].  We choose 
these parameters to reflect the results of the hydrodynamic simulations of Ref.~\cite{rudd_etal08}.

We study constraints that one would expect from three forthcoming galaxy imaging surveys.  As mentioned 
in \S~\ref{subsection:observables}, we follow current convention by setting the intrinsic shape noise 
of source galaxies to be $\gi=0.2$ for all experiments and 
subsuming differences in the shape noise between surveys into 
differences in the effective number density of galaxies on the sky for each survey.  
We assume a redshift distribution of source galaxies as given in Eq.~(\ref{eq:sources}).

\FIGURE[t!]{\epsfig{file=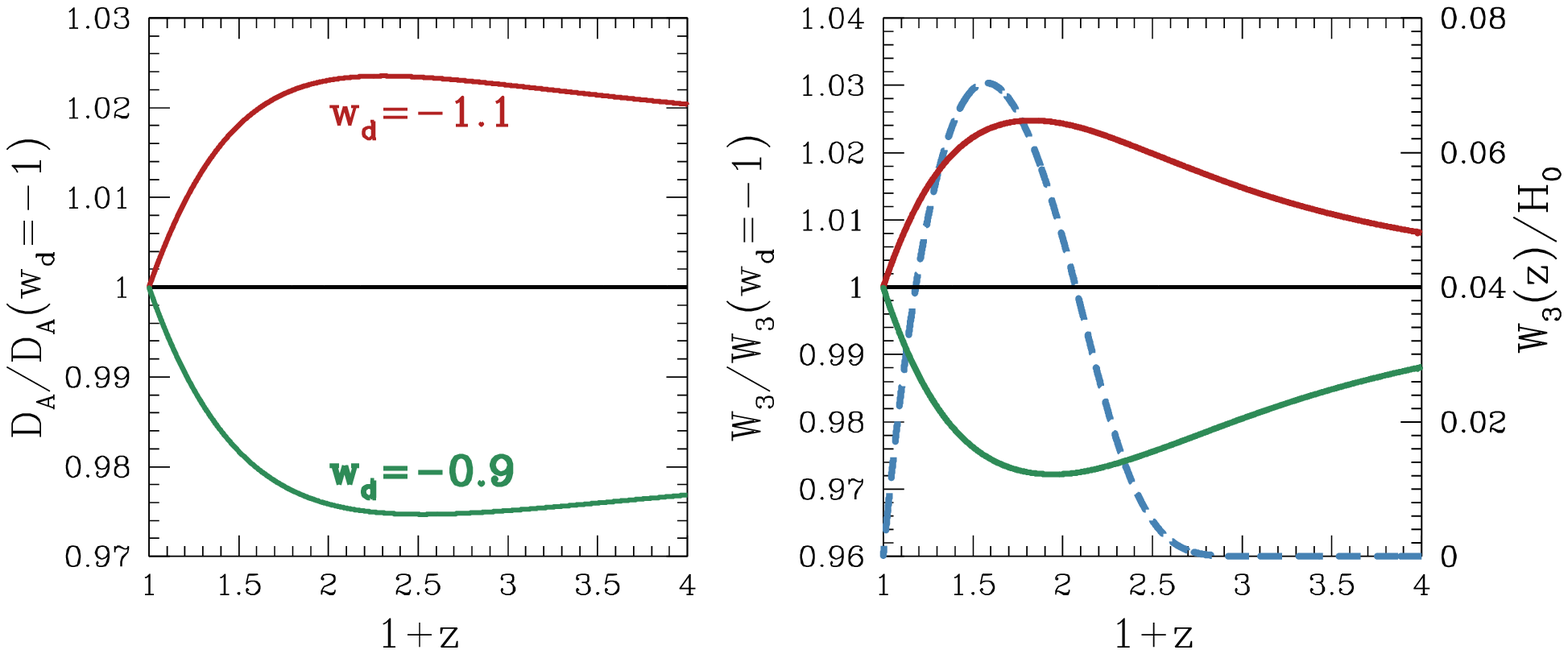,width=15cm}\caption{
The influence of the dark energy equation of state parameter on the relationship 
between angular diameter distance and redshift.  The {\em left} panel shows the angular diameter 
distance as a function of redshift for models with $\wdist \ne -1$, in units of the 
angular diameter distance in our fiducial model with $\wdist = -1$.  The {\em upper} line 
shows $\Da$ with $\wdist=-1.1$ and the {\em lower} line shows $\Da$ with $\wdist=-0.9$.  
The {\em right} panel shows the net influence of this distance change on the lensing 
kernel of Eq.~(\ref{eq:weight}).  The {\em dashed} line in this panel should be read 
against the right, vertical axis and represents the absolute weight $\wlw{3}(z)/H_0$.  
The solid lines show the particular example of the weight for the third tomographic bin 
$\wlw{3}(z)$ relative to its value in the $\wdist=-1$ case and should be read against 
the left vertical axis.  The correspondence with $\wdist$ is 
as in the {\em left} panel. 
}
\label{fig:distances}}

The most near-term survey that we explore is the dark energy survey (DES), which may begin operations 
in 2009 and have results as soon as 2011-2012\footnote{{\tt http://www.darkenergysurvey.org}}.  For the 
DES, we take $\fsky=0.12$ and $\bar{n}=15/\mathrm{arcmin}^2$.  We also consider an imaging survey 
that might be conducted as part of a future space-based mission such as the proposed 
Supernova Acceleration Probe (SNAP)\footnote{{\tt http://snap.lbl.gov}}, which is a canonical 
example of a National Aeronautics and Space Administration, Beyond Einstein, 
Joint Dark Energy Mission probe\footnote{{\tt http://universe.nasa.gov/program/probes/jdem}}, 
though the specifics of the mission have yet to be decided and other competitors include 
the Advanced Dark Energy Physics Telescope 
(ADEPT)\footnote{{\tt http://universe.nasa.gov/program/probes/adept}} 
and the Dark Energy Space Telescope 
(DESTINY)\footnote{{\tt http://www.noao.edu/noao/staff/lauer/destiny}}.  
For a SNAP-like probe, we take $\fsky=0.025$ and $\bar{n}=100/\mathrm{arcmin}^2$.  
Lastly, we consider a future ground-based imaging survey as might be carried out by the 
Large Synoptic Survey Telescope (LSST)\footnote{{\tt http://lsst.org}}.  We adopt 
$\fsky=0.5$ and $\bar{n}=50/\mathrm{arcmin}^2$ to describe the LSST survey.  In all 
cases we consider only the statistical limitations of the surveys.  The only systematic 
uncertainty we consider is the theoretical uncertainty associated with the inability to 
make precise predictions of the net influence of baryons on the lensing power spectra.

\FIGURE[t!]{\epsfig{file=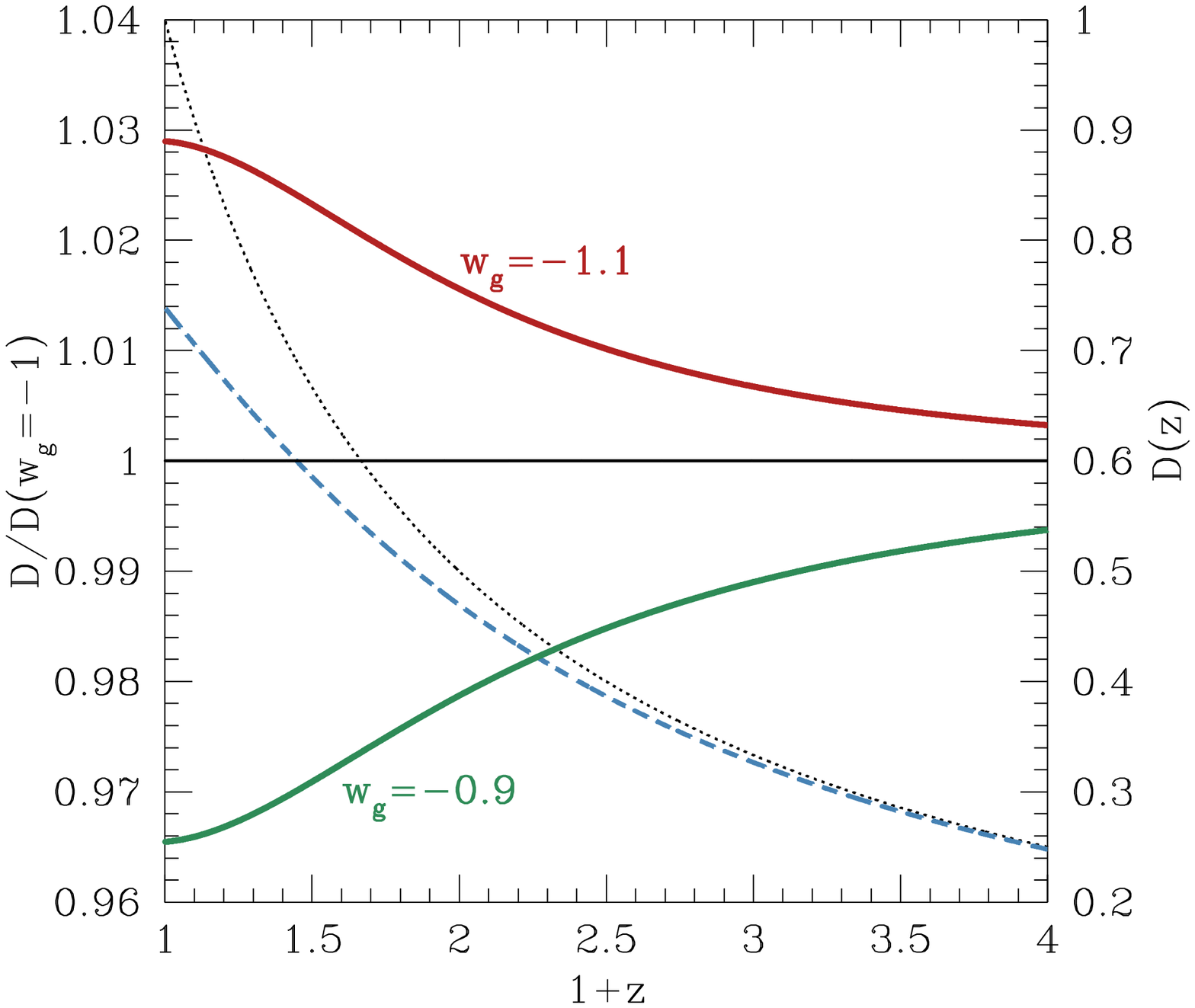,width=10cm}\caption{
The influence of the dark energy equation of state parameter on the linear 
growth of structure over an observationally-relevant range of redshifts for 
forthcoming imaging surveys.  We plot the linear growth function for dark energy 
with $\wgrow=-1.1$ ({\em upper} line) and $\wgrow=-0.9$ ({\em lower} line) 
as relative to the growth function in a cosmological constant model with 
$\wgrow=-1.0$.  These relative shifts should be read against the left, vertical 
axis.  In analogy with Fig.~\ref{fig:distances}, we plot the absolute growth function 
for the standard case of $\wgrow=-1$ as the {\em dashed} line, which should be read 
against the right, vertical axis.  The {\em dotted} line shows the growth function in 
a flat cosmological model with $\Omegam=1$ and $\Omegade=0$ (and should again be read 
against the right, vertical axis).  Comparing this growth rate to the 
growth rate in the fiducial model with $\Omegade=0.76$ illustrates the suppression 
of growth caused by the contemporary epoch of accelerated expansion.
}
\label{fig:growth}}

\section{RESULTS}
\label{section:results}

\subsection{Consistency Checks}
\label{sub:consistency}

We begin by illustrating the power of weak lensing surveys to test the consistency 
of general relativity with forthcoming data.  As we have mentioned above, the consistency 
checks work by comparing simultaneous constraints on cosmological distances and 
structure growth and determining whether they are consistent with a single, underlying, 
general relativistic model.  Before moving on to parameter constraints we briefly 
discuss each of these effects individually.  We 
will use the split dark energy parameterization model (with parameters 
$\wgrow$ and $\wdist$) to illustrate the influence of modified distances 
and modified growth.  Those readers with significant experience with dark 
energy phenomenology and proposed consistency checks of general relativity 
may like to proceed to the next section.

The influence of dark energy on the cosmic distance scale is demonstrated 
in Figure~\ref{fig:distances}.  The left panel of this figure shows the angular 
diameter distance as a function of redshift in dark energy models with constant 
equations of state $\wdist=-1.1$, $\wdist=-1.0$, and $\wdist=-0.9$.  The more 
negative the equation of state parameter, the more recent the prevalence of dark 
energy, and the more rapid is the current acceleration.  Thus, for a fixed present Hubble 
expansion rate, lower values of $\wdist$ lead to larger 
angular diameter distances.  The dependence of the angular diameter distance on the dark 
energy equation of state manifests in the lensing weight [Eq.~(\ref{eq:weight})]; this 
dependence implies that the lensing weight varies with $\wdist$.  The effects of this dependence are shown in the right panel of 
Fig.~\ref{fig:distances} for the particular case of the weight for the third 
tomographic bin containing $\wlw{3}(z)$, which contains sources with photometric 
redshifts in the range $1.2 \le z_p < 1.8$.  Notice that the lens weight itself 
(dashed line) extends to $ z>1.8 $.  In our model, sources in this bin have true redshifts 
that extend beyond $z=1.8$ because of the relatively large 
dispersion in calibrated photometric redshifts.  For more negative $\wdist$, the $\wlw{3}(z)$ 
is relatively increased and weighted relatively more toward low redshift.  Heuristically, 
this can be understood by considering a hypothetical individual deflector.  Relative 
to the deflector, the angle of deflection of a light ray is fixed by the deflector properties 
and the apparent change of position of the source as seen by an observer grows with angular 
diameter distance.  The varying distances to fixed redshift also affect the mapping between multipole and 
wavenumber in the matter power spectrum at fixed redshift.  Increased distances 
mean smaller wavenumber at fixed redshift, and as the power spectrum of density fluctuations 
is a declining function of wavenumber on scales of interest, this results in slightly 
stronger lensing.

\FIGURE[t!]{\epsfig{file=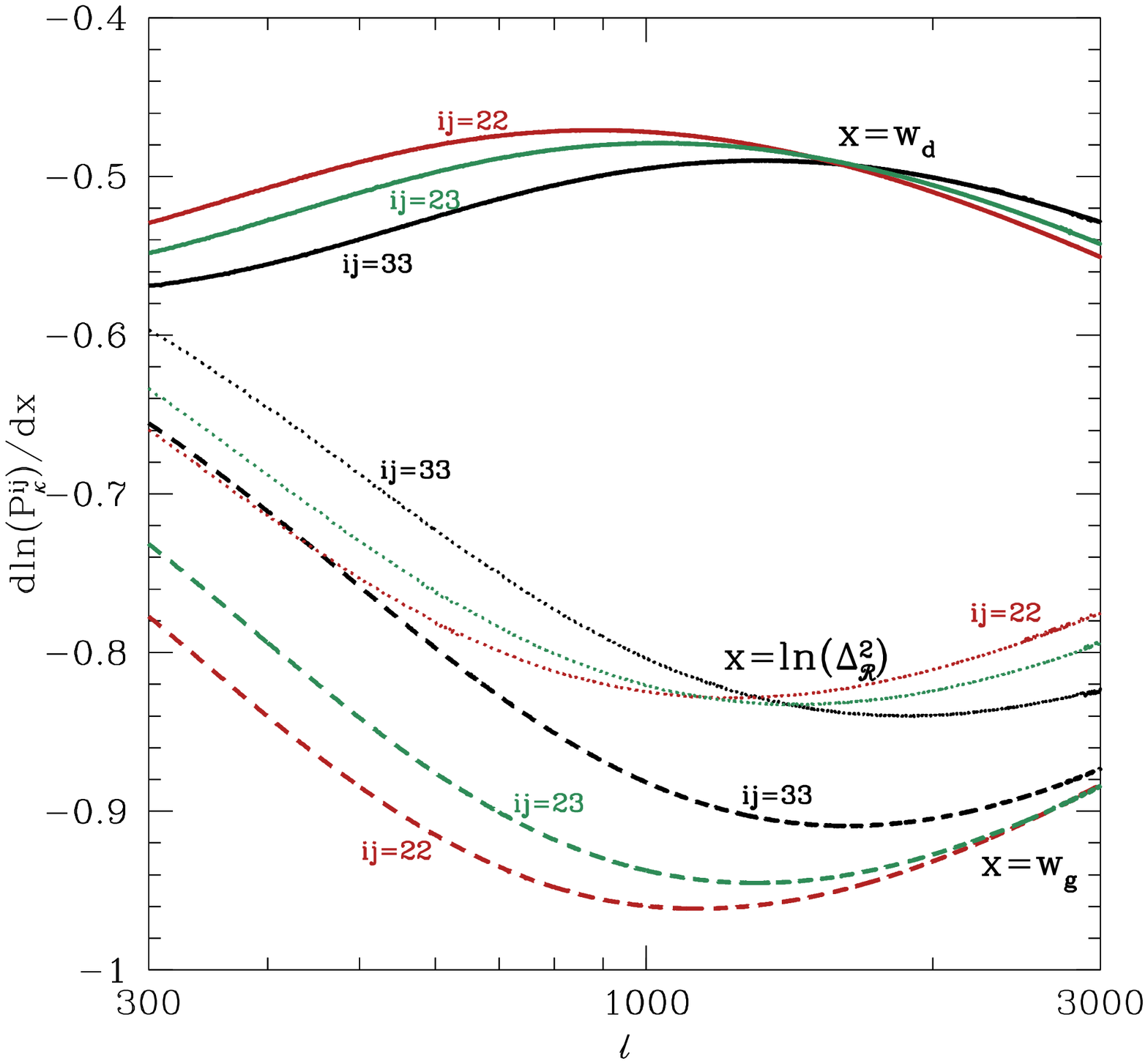,width=12cm}\caption{
Partial derivatives of tomographic weak lensing power spectra with respect to 
cosmological parameters.  We show derivatives of each of the power spectra 
$\Pktom{i}{j}(\ell)$, with $ij=22,23,33$ with respect to the three parameters 
$\wdist$, $\wgrow$, and $\ln \dr$.  The {\em solid} lines show derivatives 
with respect to $\wdist$.  From bottom to top at left, the {\em solid} lines are 
$\partial \ln \Pktom{i}{j}/\partial \wdist$ with $ij=33,23,22$.  The {\em dashed} lines 
represent derivative with respect to $\wgrow$.  From bottom to top at left, 
the {\em dashed} lines show $\partial \ln \Pktom{i}{j}/\partial \wgrow$ with 
$ij=22,23,33$.  The {\em dotted} lines are derivatives with respect to the power 
spectrum normalization parameter $\ln \dr$.  From bottom to top at left, the 
{\em dotted} lines show $-\partial \ln \Pktom{i}{j}/\partial \ln \dr$ with 
$ij=22,23,33$.  The additional negative sign for the derivatives with respect to 
$\ln \dr$ is designed to reduce the dynamic range on the vertical axis.
}
\label{fig:wderivs}}

Next, we demonstrate the dependence of the cosmic growth function on dark energy.  
Figure~\ref{fig:growth} shows the growth function $D(z)$ in our fiducial cosmology 
(dashed line), normalized such that $D(z) \rightarrow (1+z)^{-1}$ as $z\rightarrow \infty$.  
The dashed line in this figure shows the growth function in a flat cosmological model 
with no dark energy and $\Omegam=1$.  Comparing these two growth functions shows the 
dramatic suppression in the growth of density perturbations during the epochs where dark 
energy makes a significant contribution to the cosmic energy budget.  
In addition, we show the growth functions in cosmological models with $\wgrow=-1.1$ and 
$\wgrow=-0.9$ relative to the growth function in our fiducial model.  The more positive 
the dark energy equation of state is, the earlier the acceleration begins and the earlier 
the growth of structure is quenched by the competing cosmic expansion.  Conversely, more negative 
values of the dark energy equation of state lead to more recent expansion and, as a result, 
more aggregate growth of density perturbations since the epoch of recombination.  These 
features are all represented in Fig.~\ref{fig:growth}.

The modifications to the growth of structure and cosmic distances manifest as 
changes in the observable power spectra.  Figure~\ref{fig:wderivs} shows partial derivatives 
of three of the power spectra, $\Pktom{2}{2}$, $\Pktom{3}{3}$, and $\Pktom{2}{3}$, 
as a function of parameters $\wdist$, $\wgrow$, and $\ln \dr$.  The derivatives 
with respect to the dark energy parameters are all negative.  To reduce the range of 
the vertical axis, we actually display $-\dd \Pktom{i}{j}/\dd \ln \dr$, so that it will 
also lie in the negative vertical half plane.  Notice that variations in the observable 
power spectra due to changes in $\wgrow$ and $\wdist$ 
have very different redshift and scale dependence.  Consequently, we should not 
expect these parameters to exhibit a significant degeneracy and we would expect that 
shear power spectrum observations could constrain both parameters independently.  
Notice also that changes in $\wgrow$ induce relatively larger shifts in the observable 
power spectra than do shifts in $\wdist$.  In the absence of additional information, 
this might be taken as evidence that weak lensing constraints on $\wgrow$ should be 
more stringent than such constraints on $\wdist$.  In practice, $\wgrow$ is strongly 
degenerate with the power spectrum normalization parameter $\ln \dr$.  This can also be 
seen in Fig.~\ref{fig:wderivs}.  The dotted lines 
in Fig.~\ref{fig:wderivs} shows $-\dd \Pktom{i}{j}/\dd \ln \dr$, and this quantity 
exhibits a similar redshift and scale dependence as the derivatives of the spectra 
with respect to $\wgrow$.  Though not shown, $\wgrow$ is also strongly degenerate 
with $\omegab$.  The net result of these degeneracies is that $\wdist$ will be 
significantly more strongly constrained by weak lensing power spectrum observations 
than $\wgrow$.  This discussion is closely related to the eigenmode analysis of 
Ref.~\cite{zhan_etal08} and serves as a qualitative demonstration that both $\wgrow$ 
and $\wdist$ can be constrained by cosmic shear measurements, but the relative constraints 
on each parameter are sensitive to choices for external priors in the available additional parameter 
space.

Having reviewed the influence of independent changes in the cosmological distance scale 
and the rate of density fluctuation growth on observable weak lensing power spectra, 
we now turn to projections for the utility of general relativity consistency checks.  
We utilized the split dark energy equation of state parameterization for the 
previous illustrations, but we present results for both the $\wdist$-$\wgrow$ 
parameterization as well as the growth index ($\gamma$) parameterization.  
Standard projections for the power of forthcoming consistency checks using weak 
lensing are summarized in Figure~\ref{fig:consist}.  Clearly such consistency checks can be 
quite powerful when definitive predictions can be made for the convergence power 
spectra.  In the particular case of LSST, projected $1\sigma$ 
constraint on $\wdist$ is roughly $\sigma(\wdist) \simeq 0.02$, while 
the projected constraint on the gravitational growth index is 
$\sigma(\gamma) \simeq 0.04$, meaning that the LSST weak lensing program 
alone could rule out DGP gravity at the $\sim 3.5\sigma$ level.  In the case of the 
$\wdist$-$\wgrow$ parameterization, it is the equality of these two parameters that 
serves as a null hypothesis, so it is useful to look at the marginalized constraint 
on the combination $\wdiff \equiv \wdist-\wgrow$.  Transforming to this variable, 
we find a $1\sigma$ constraint $\sigma(\wdiff)=0.04$ for LSST, $\sigma(\wdiff)=0.11$ 
for SNAP, and $\sigma(\wdiff)=0.14$ for DES.

Unfortunately, these standard projections assume perfect knowledge of nonlinear 
structure growth and ignore the influence of poorly-understood baryonic processes on 
the predictions of lensing spectra.  Uncertainties due to baryonic processes and 
the process of galaxy formation are significant and will be the subject of the remainder 
of this section, beginning with the following section where we demonstrate that 
such effects can significantly bias estimators of dark energy parameters and diminish 
the utility of consistency checks if not accounted for.

\FIGURE[t!]{\epsfig{file=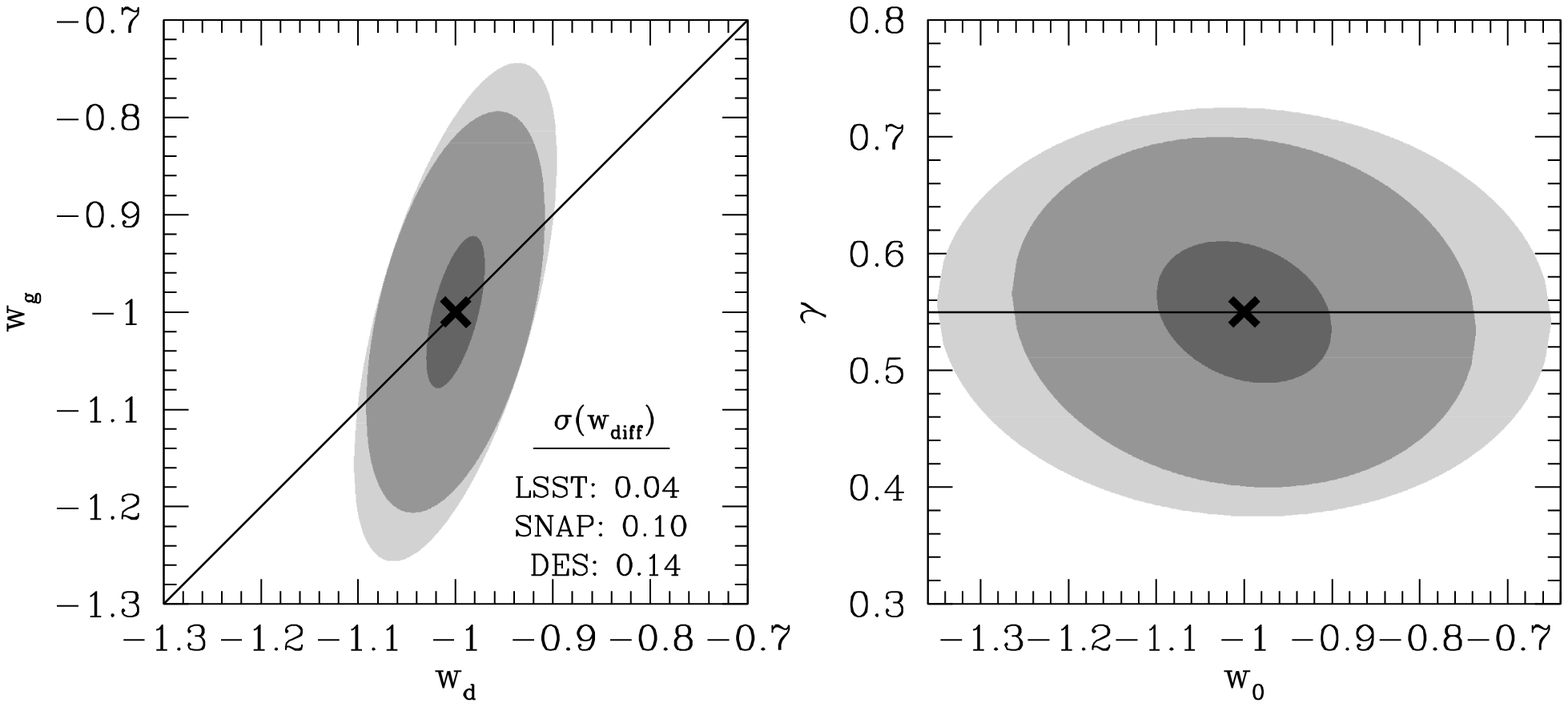,width=15cm}\caption{
The capability of forthcoming surveys to perform consistency 
checks of general relativity.  The {\em left} panel shows projected 
$1\sigma$ constraint contours in the $\wdist$-$\wgrow$ plane for 
the LSST (innermost contour), DES (outermost contour), and 
SNAP weak lensing (middle contour) experiments.  The diagonal line 
in this panel delineates $\wgrow=\wdist$.  The constraints on the 
parameter $\wdiff \equiv \wdist-\wgrow$ from each experiment are 
given in the {\em lower, right} portion of the panel.  The {\em right} panel 
shows constraints in the gravitational growth index parameterization 
in the $\wzero$-$\gamma$ plane.  The three contours represent the 
LSST, DES, and SNAP weak lensing experiments as in the {\em left} panel.  
}
\label{fig:consist}}

\subsection{Biases}
\label{sub:biases}

The poorly-constrained baryonic processes are detrimental to any consistency check of 
general relativity because, if unaccounted for, they may induce biases in inferred 
cosmological parameters.  In particular, if the estimators of dark energy parameters that 
we have introduced are significantly biased, it would be possible to conclude erroneously 
that data are inconsistent with general relativity.  The relevant quantities to 
examine in order to assess the importance of baryonic effects are the biases 
in units of the statistical uncertainties of those parameters.  If the biases are small 
compared to the statistical uncertainties, it will be unlikely to rule out the true model 
based on these biases, but as biases become comparable to or larger than statistical 
uncertainties it becomes increasingly likely that the true model may be ruled out based 
on the data.

\FIGURE[th!]{\epsfig{file=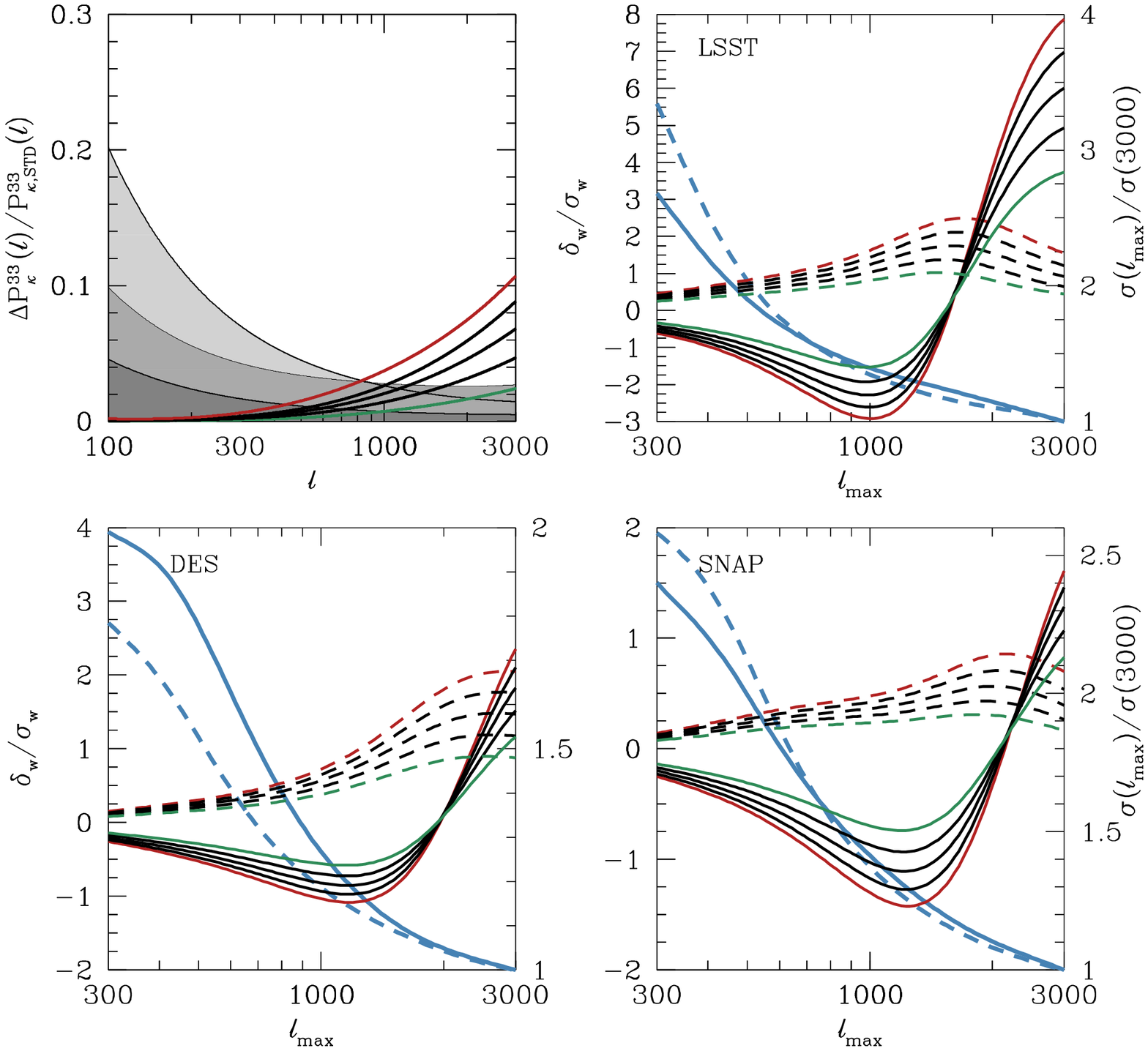,width=12.5cm}\caption{
Biases in the estimators for the split dark energy parameters 
$\wgrow$ and $\wdist$ that may be realized if baryonic 
processes are ignored.  The {\em upper, left} panel shows the the effect 
of modified halo structure on convergence power spectra.  
Each of the lines that increases with multipole 
represents the relative change in a convergence power spectrum of sources in 
our third tomographic bin ($1.2 \le z_p < 1.8$) $\Pktom{3}{3}(\ell)$, in 
models with modified halo structure relative to that of 
the standard case.  We represent the standard case using a halo model 
where halo concentrations are given by Eq.~(\ref{eq:cofm}) with 
$c_0=10$, $\alpha=0.1$, and $\beta=1.0$.  We represent models with modified 
halo structure by taking $c_0=11,12,13,14,\mathrm{\ and,\ }15$ from bottom to top.  
The shaded bands show the statistical errors on $\Pktom{3}{3}(\ell)$ expected from 
forthcoming SNAP, DES, and LSST surveys from {\em top to bottom at left}.  The other 
three panels show the biases in estimators $\wdist$ ({\em solid}) and $\wgrow$ ({\em dashed}), in 
units of the statistical uncertainties in these parameters, as a function 
of the maximum multipole used in parameter estimation.  These should be read against the 
{\em left} vertical axes.  Each panel shows forecasts for a specific survey.  The most 
biased cases correspond to $c_0=15$.  The decreasing functions of $\ellmax$ in 
each panel show the statistical uncertainty in $\wdist$ and $\wgrow$ as a function of $\ellmax$ 
relative to the error if all information to $\ellmax=3000$ were used.  These lines should be 
read against the {\em right} vertical axes.
}
\label{fig:wbiases}}

In practice it is difficult to estimate what biases may be realized, 
because the true power spectra must be known in order to perform 
such a calculation (and, of course, if the true spectra were calculable there would be no bias).  
Fig.~\ref{fig:wbiases} gives an estimate of biases that may reasonably be realized for a variety 
of assumptions regarding the true convergence power spectra.  We constructed these bias estimates 
as follows.  Following the simulation analysis of Refs.~\cite{zentner_etal08}, we 
assume that the primary influence of baryons can be accounted for in a halo model by adopting 
a non-standard halo concentration relation.  It has been demonstrated that incorporating this 
additional freedom into the halo model enables accurate modeling of the results of non-radiative 
and dissipational hydrodynamic cosmological simulations and reduces expected biases in inferred 
dark energy parameters to acceptable levels \cite{zentner_etal08}.  The standard result for the concentration 
relation from gravity-only simulations is Eq.~(\ref{eq:cofm}) with roughly $c_0=10$, $\alpha=0.1$, 
and $\beta=1.0$ \cite{neto_etal07}.  We take this model to be our treatment of the results of 
$N$-body simulations.  The halos in the simulations of Ref.~\cite{rudd_etal08} exhibit a $\sim 40\%$ 
enhancement in their concentrations in dissipational simulations relative to $N$-body simulations, so a 
choice of $c_0 \approx 14$ is an acceptable model for these simulation results.  Unfortunately, our 
understanding of the evolution of baryons is poor and no simulation campaign can produce definitive 
results (it is this lack of definitiveness that is the primary problem).  In particular, this boost 
is likely an over-estimate of any true shift due to galaxy formation because contemporary simulations 
exhibit baryonic cooling and star formation that is more efficient than allowed observationally 
\cite{katz_white93,lewis_etal00,pearce_etal00,balogh_etal01,suginohara_ostriker98}.  As a consequence, 
baryonic processes must be treated in a way that accommodates a wide range of predictions.  We illustrate 
reasonable levels of bias due to galaxy formation effects by computing the biases induced in dark energy 
parameter estimators when the convergence spectra are modeled with a fixed halo concentration-mass 
relation with $c_0=10$, but where the ``true'' value of this normalization ranges among 
$c_0=11,12,13,14,\mathrm{\ and,\ }15$.  The lower end of this range is only a small 
shift relative to the standard concentration relation and is realized even in non-radiative 
hydrodynamic simulations where baryons cannot radiate their energy and form galaxies.  The upper 
end of this range is slightly larger than the Ref.~\cite{rudd_etal08} simulation results 
and is near the maximum allowable by current lensing constraints at low redshift ($z \sim 0.2$) 
\cite{mandelbaum_etal08}.  We note in passing that there are as yet weak observational indications 
of concentrations higher than those predicted by dissipationless simulations 
\cite{buote_etal06,vikhlinin_etal06}.

Fig.~\ref{fig:wbiases} shows biases in the inferred $\wgrow$ and $\wdist$ induced by treating 
convergence power spectra that would be well described by an enhanced concentration 
mass relation, where $c_0$ can be any of $c_0=11,12,13,14,\mathrm{\ or\ }15$, with a 
concentration-mass relation fixed to have $c_0=10$.  The objective is 
to provide reasonable estimates for biases that may be realized by not 
treating baryonic processes appropriately.  If the Ref.~\cite{rudd_etal08} simulation 
results were a good representation of the ``true'' convergence power spectra, the 
biases would be near the top of the range in Fig.~\ref{fig:wbiases}, roughly corresponding 
to the fourth most strongly biased model in each panel.

To see why it is useful to consider biases in dark energy 
parameter estimators as a function of scale may require some elaboration.  
The influence of baryons is greater on smaller scales, 
so a simple way to eliminate the uncertainty caused by baryons is 
to excise small-scale information from parameter estimation.  A simplistic way to do this 
is to choose a maximum multipole $\ellmax$ for the analysis and to disregard spectra beyond that maximum 
(Ref.~\cite{huterer_white05} explores more sophisticated means to excise small-scale 
information, but they yield similar results in this context \cite{zentner_etal08}). 
In Fig.~\ref{fig:wbiases}, we show biases as a function of $\ellmax$ in order 
to illustrate the reduced biases realized as a function of minimum scale (maximum-multipole).

Fig.~\ref{fig:wbiases} illustrates several points of interest.  Most previous lensing 
analysis have assumed that one could utilize information to at least $\ellmax \sim 10^3$ 
effectively.  Fig.~\ref{fig:wbiases} shows that if one considers such small-scale information, 
estimators of dark energy equation of state parameters $\wgrow$ and $\wdist$ will both be significantly 
biased relative to their true values.  Notice also that they are generally biased in an 
{\em opposing} sense.  This is of paramount importance for consistency checks where the goal 
is not necessarily to constrain either parameter individually, but to test the hypothesis that 
$\wgrow$ and $\wdist$ are the same.  For example, if $\wgrow$ and $\wdist$ were always biased 
in the same way due to an unaccounted for systematic then it would not be possible to produce a reliable 
constraint on either parameter, but in this contrived example it would still be possible to test for 
their equality without accounting for the systematic.  The biases in Fig.~\ref{fig:wbiases} are generally 
large compared to statistical uncertainties and suggest that consistency checks of general relativity 
can be compromised by uncertainty in galaxy formation physics and may lead to rejection of the 
hypothesis of general relativity even in models where this is the true theory of gravity.  
The point where biases become acceptable ($\delta_w \ll \sigma_w$)
varies from one experiment to the next.  One could argue that for the LSST survey these biases 
are not acceptable for any $\ellmax \gtrsim 300$.

\FIGURE[th!]{\epsfig{file=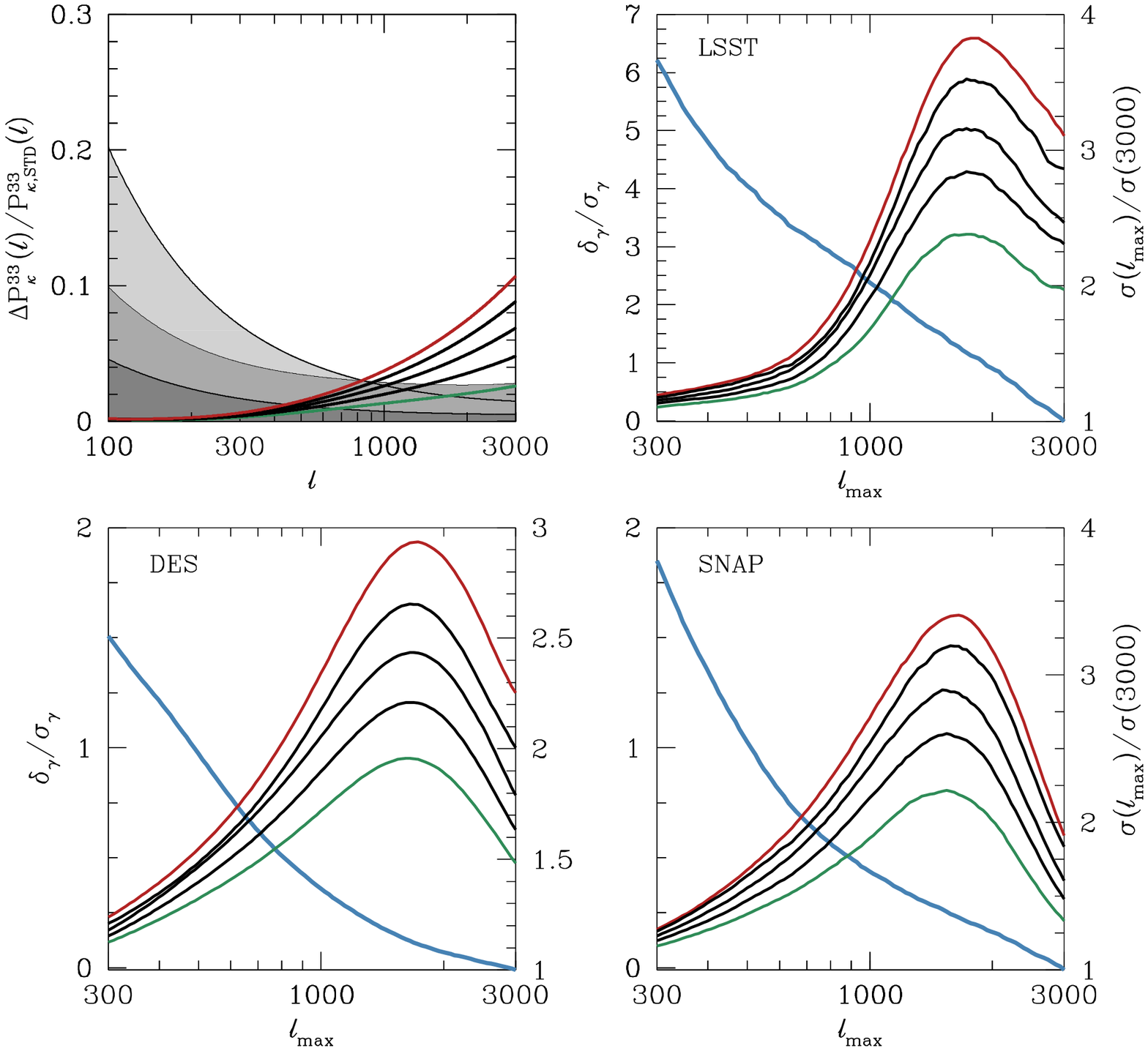,width=12.5cm}\caption{
Estimates of biases in the estimator of the $\gamma$ parameter of the gravitational 
growth index formalism proposed in Ref.~\cite{linder05}.  The treatment of baryonic 
modifications to the convergence power spectrum is as in Fig.~\ref{fig:wbiases} and the 
{\em upper, left} panel is identical to the {\em upper, left} panel of Fig.~\ref{fig:wbiases}.  
We repeat it here for convenience.  The remaining panels show biases projected 
for the LSST, SNAP, and DES weak lensing surveys as indicated.  Value of the biases should 
be read relative to the {\em left} vertical axes.  The biases are given in 
units of the projected statistical uncertainty in $\gamma$ and the most (least) biased 
case corresponds to a model with $c_0=15$ ($c_0=11$).  The intermediate lines correspond 
to the intermediate values of $c_0$ with bias increasing with $c_0$.   The monotonically 
decreasing lines in each panel represent the $1\sigma$ statistical uncertainties in 
$\gamma$ as a function of maximum multipole $\ellmax$.  These are shown in units of the 
uncertainties in gamma achieved by considering all information to $\ellmax=3000$ in order 
to illustrate the relative degradation in constraints caused by excising small-scale 
information.  
}
\label{fig:gbiases}}

As a rough criterion, Fig.~\ref{fig:wbiases} demonstrates that reducing the biases induced 
by the shifts in halo structure to acceptable levels 
requires taking a maximum multipole of no more than a few hundred.  
Of course, this reduction in information comes at a cost.  The monotonically 
decreasing lines in Fig.~\ref{fig:wbiases} show the marginalized $1\sigma$ 
statistical uncertainties in $\wdist$ and $\wgrow$ as a function of the maximum 
multipole $\ellmax$ in units of the uncertainty when all information to an $\ellmax=3000$ 
is used.  Notice that these uncertainties should be read relative to the right vertical 
axes in each panel.  Overlaying the dependence of parameter uncertainties on this plot 
serves to demonstrate the cost of excising small-scale information.  For LSST, the cost is 
large, a factor of $\sim 3-4$ in the uncertainties of both $\wdist$ and $\wgrow$, because of 
the exquisite precision with which LSST could measure convergence spectra.  This greatly 
reduces the effectiveness of an LSST-like data set to test the consistency of general 
relativity.  DES is least affected by the uncertainty in galaxy formation because it 
surveys a comparably low number density of galaxies and makes the least use of small-scale 
information of any of the three experiments.  SNAP is intermediate between the two with 
constraints degraded by roughly a factor of $\sim 2.5$.

Figure~\ref{fig:gbiases} shows the biased estimator for the gravitational growth index 
introduced in Ref.~\cite{linder05}.  Our methods for computing biases and the 
presentation are analogous to that of Fig.~\ref{fig:wbiases}.  Of course, 
in this case the only parameter of importance is the gravitational growth 
index $\gamma$.  Qualitatively, Fig.~\ref{fig:gbiases} shows results that 
are similar to that of Fig.~\ref{fig:wbiases}.  In particular, biases can 
be considerable compared to statistical uncertainties.  Moreover, excising 
small-scale information comes at great cost in the statistical uncertainties 
of the dark energy parameters, with relative degradation of $\sim 2-4$ 
depending upon the choice of $\ellmax$ and experimental parameters.

\subsection{Calibration}
\label{sub:calibration}

In the previous section, we demonstrated the significance of the biases that would be induced 
in tests of general relativity if baryonic processes were to be ignored.  We also showed the 
significant cost of excising information from the small scales where baryonic effects are most 
important in terms of degraded parameter uncertainties and thus degraded tests of the 
theory of gravity.  What remains is to study whether such tests can still be performed utilizing 
high-multipole information if baryonic effects are treated in a parameterized way and calibrated 
simultaneously with the cosmological parameters.  Though there is still significant work to be done, 
preliminary indications are that baryonic effects can indeed be treated by modifying the predictions 
of dissipationless $N$-body simulations \cite{rudd_etal08,rudd_etal08}.  In fact, the necessary modification is 
to treat the relation between halo mass and halo concentration as free, and we have already utilized such 
modifications to estimate parameter biases in the previous section.

The next step is to determine the degree to which treating both halo structure and cosmology as uncertain, 
in order to eliminate the biases of the previous section, and fitting both to forthcoming data will 
degrade our ability to test general relativity.  If the degradation is significantly less than that 
incurred by simply excising small-scale information, then an appropriate strategy to adopt in the run-up 
to the next generation of imaging surveys would be to spend significant effort understanding the 
phenomenology of structure growth in the nonlinear regime in a variety of different effective models 
for the evolution of the baryonic sector.  It might then be possible to ``self-calibrate'' the 
influence of baryons, simultaneously yielding insight into the law of gravity on large scales 
and the small-scale physics of galaxy formation.  Alternatively, if the degradation is comparable to the 
degradation incurred by excising small-scale information, such a program would likely be futile and 
it would be wiser to excise small-scale information until such time as definitive predictions for 
the influence of baryonic processes on lensing power spectra can be made.

There is already reason 
to be optimistic that some self-calibration can be achieved.  Ref.~\cite{huterer_etal06} and 
Ref.~\cite{albrecht_etal06} demonstrated the robustness of weak lensing to modest multiplicative 
and additive errors in shear measurements.  In addition, Ref.~\cite{zentner_etal08} 
have already demonstrated robustness of weak lensing constraints on dark energy to some 
uncertainty in the power spectrum at high wavenumbers.  A large part of potential to carry 
out self-calibration programs within weak lensing data can be gleaned from the eigenmode analysis of 
Ref.~\cite{zhan_etal08}.  The robustness of weak lensing constraints on dark energy is due in part to 
the fact that weak lensing provides exquisite constraints on distance measures that are relatively 
insensitive to dark energy and can be brought to bear on other systematics without degrading dark 
energy constraints.

In this section, we present the results of a self-calibration exercise where we treat cosmological 
parameters as before and introduce additional parameter freedom to describe the influence of baryons.  
To be specific, we treat the halo concentrations according to Eq.~(\ref{eq:cofm}), but allow the 
parameters $c_0$, $\alpha$, and $\beta$ to be free.  It is already possible to offer a guess at 
the promise of such a self-calibration exercise.  Fig.~\ref{fig:wderivs} shows the derivatives of 
three of our convergence power spectra $\Pktom{2}{2}$, $\Pktom{2}{3}$, and $\Pktom{3}{3}$ with respect 
to the parameters $\wdist$, $\wgrow$, and $\ln \dr$.  Compare this to the shifts in $\Pktom{3}{3}$ 
depicted in the upper, left panels of Fig.~\ref{fig:wbiases}.  Modifying the concentration relation 
causes significant increases in power at small scales (high multipoles).  As a result, there is a 
stronger scale dependence associated with changes in concentration parameters than there is 
associated with changes to $\wdist$, $\wgrow$, $\gamma$, or $\dr$.  This is encouraging because it 
suggests that if the scale dependence can be adequately modeled, the parameters describing baryonic 
processes can be extracted independently without significantly degrading consistency checks of 
general relativity.

\FIGURE[t!]{\epsfig{file=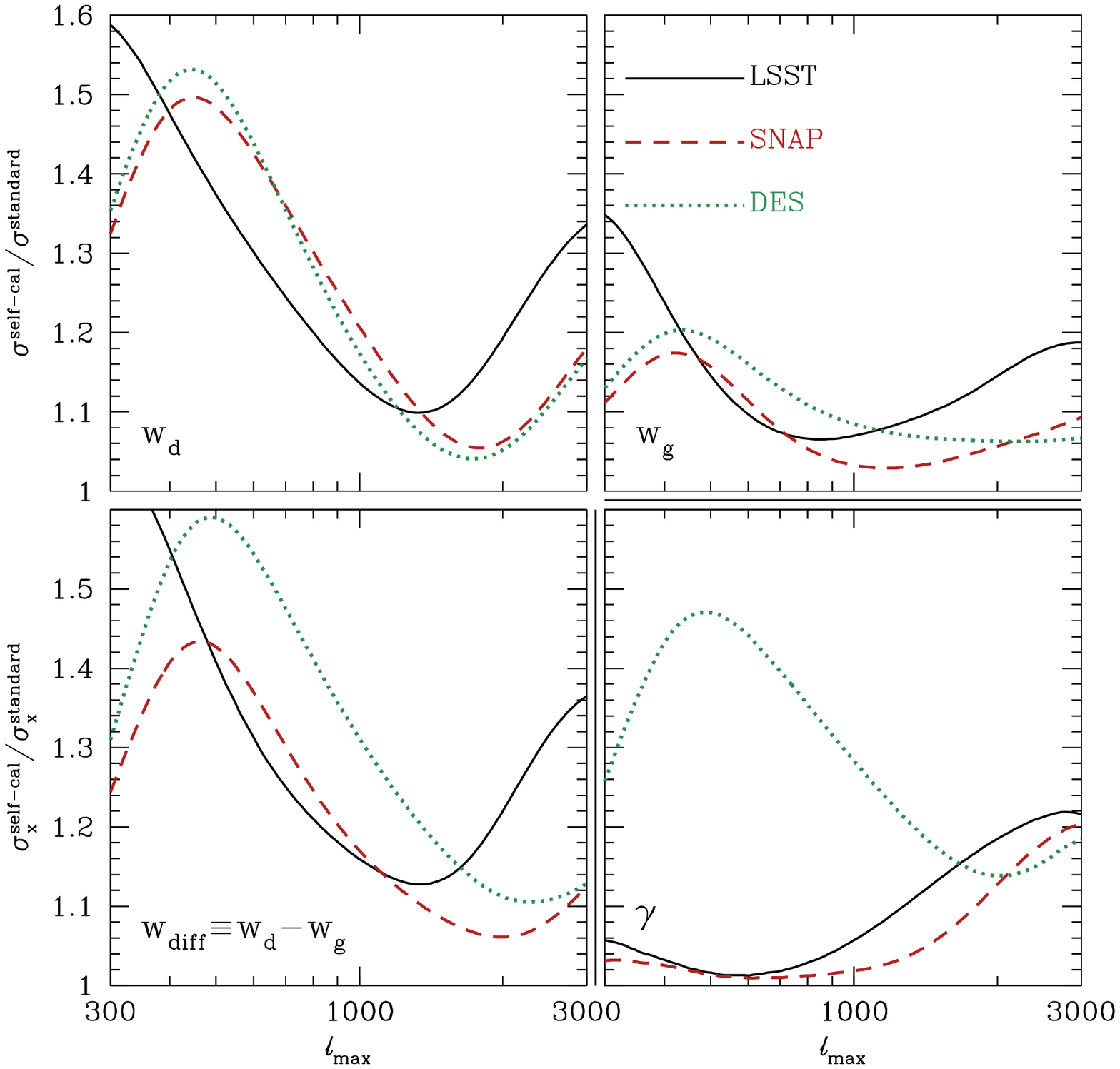,width=12cm}\caption{
Results of self-calibrating halo structure simultaneously with tests for modified gravity 
on large scales.  As with Fig.~\ref{fig:wbiases}, we present results as a function of 
the maximum multipole considered in the parameter extraction analysis $\ellmax$.  The 
vertical axis is the error on the stated model parameter in the self-calibration case 
in units of $\sigma_{X}^{standard}$ : the error on parameter X calculated by standard parameter forecasts that appear in 
the literature, where halo structure is assumed to be known perfectly.  Each panel contains 
three lines for each set of experimental parameters we explore.  The {\em solid} lines correspond 
to an LSST-like experiment, the {\em dashed} lines correspond to a SNAP-like experiment, and the 
{\em dotted} lines correspond to a DES-like experiment.  Each panel focuses on a different parameter.  
The {\em upper, left} panel shows $\wdist$, the {\em upper, right} panel shows $\wgrow$, the 
{\em bottom, left} panel shows the difference $\wdiff=\wdist-\wgrow$, and the 
{\rm bottom, right} panel shows the gravitational growth index $\gamma$.  Note that 
$\wdist$ and $\wgrow$ are the parameters of our ``dark energy split'' model and are constrained 
simultaneously, and that the relevant constraint for tests of gravity is the constraint on $\wdiff$.  
The $\gamma$ parameter is from an entirely distinct parameterization in which dark energy is 
independently marginalized over.  The additional lines that separate this panel from the others 
are intended to reinforce this distinction.
}
\label{fig:cal}}

We show the results of the self-calibration exercise in Figure~\ref{fig:cal}.  In this panel 
we show constraints in the case of self calibration of halo structure 
$\sigma^{\mathrm{self-cal}}(\ellmax)$, 
in units of the parameter uncertainties that would be quoted in the standard case of perfect 
knowledge of halo structure, $\sigma^{\mathrm{standard}}(\ellmax)$ (shown in Fig.~\ref{fig:consist}, 
Fig.~\ref{fig:wbiases}, and Fig.~\ref{fig:gbiases}).  
In particular, we show how the parameter degradation varies with $\ellmax$ because it is both 
instructive and it illustrates the balance between excising data and self-calibrating.  
Figure~\ref{fig:cal} shows results for both the $\wdist$-$\wgrow$ parameterization as well as the 
gravitational growth index ($\gamma$) parameterization.

It may seem odd that the functions in Fig.~\ref{fig:cal} do not decrease monotonically with 
increasing $\ellmax$.  These relations are not monotonic functions of $\ellmax$ because they represent 
the ratio of the error realized in the self-calibration calculation compared to the errors computed 
in the limit of perfect knowledge of the influence of baryons.  From Fig.~\ref{fig:wbiases} and 
Fig.~\ref{fig:gbiases}, the constraints on all parameters are rapidly decreasing functions of 
$\ellmax$.  This is sensible; as more information is added constraints should improve and not 
degrade.  Likewise, in the case of self-calibration the absolute constraints on each parameter 
decrease rapidly with increasing $\ellmax$.  What is not monotonic is the relative degradation 
of these constraints as a function of $\ellmax$.

The scale dependence of the parameter degradation is not surprising.  From the power spectra 
in the upper, left panel of Fig.~\ref{fig:wbiases} and the derivatives in Fig.~\ref{fig:wderivs} it 
is evident that each of the parameters of interest induces a scale-dependence on the observed 
spectra, so some scales are more effective at constraining dark energy parameters and calibrating 
halo structure than others.  Though the details depend upon the experiment, and in particular the 
statistical weight an experiment places on a particular multipole, there are sensible general 
trends depicted in Fig.~\ref{fig:cal}.  In the limit of low-$\ell$, halo structure is unimportant 
and results in the self-calibration calculation approach the standard results.  This is least 
evident for LSST because LSST has sufficient $\fsky$ to make precise power spectrum measurements 
even for $\ell \sim 300$.  We do not extend these plots to $\ell < 300$ because at this value of 
$\ell$ parameter biases are already at or near acceptable levels and so there is no need to 
relegate ourselves to such low multipoles in any data analysis.  With increasing multipole, 
the halo structure becomes ever more important and is mildly degenerate with dark energy.  
Further increasing $\ellmax$ beyond $\sim$~ several hundred introduces the strong scale dependence 
that halo structure imparts on the convergence spectra and this breaks a degeneracy between 
dark energy and halo structure.  For this reason, the degradation decreases rapidly from 
$\ellmax \sim 500$ to $\ellmax \sim 2000$.

\FIGURE[t!]{\epsfig{file=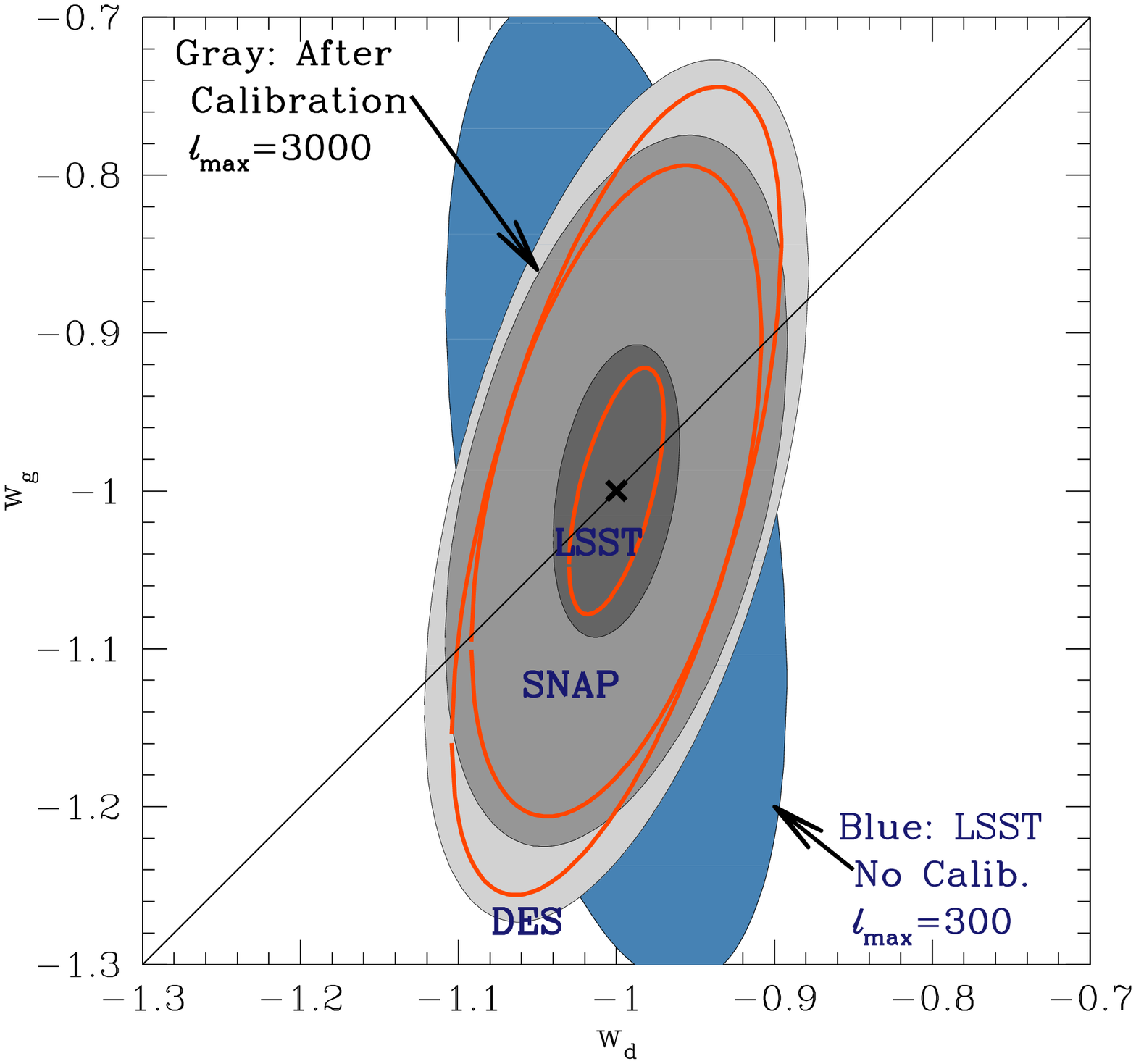,width=10cm}\caption{
Constraint contours in the $\wdist$-$\wgrow$ plane after accounting for and marginalizing over 
the uncertainty in halo structure.  The {\em filled, grey} contours correspond to the 
$1\sigma$ contours in this plane from the LSST-like (innermost contour), 
SNAP-like (middle contour), and DES-like (outermost blue contour) experiments we consider.  
The {\em thick, solid} lines overlaying the contours represent the confidence contours in the 
limit of perfect knowledge of nonlinear structure formation as shown in the left panel of 
Fig.~\ref{fig:consist}.  The cross shows the fiducial model and the diagonal line running 
from the lower left to the upper 
right delineates $\wdist$=$\wgrow$.  As an explicit demonstration of the effect of excising small-scale 
information to eliminate potential bias, the {\em outermost, blue} contour shows the $1\sigma$ 
marginalized constraint in the $\wdist$-$\wgrow$ plane from LSST if no information from multipoles 
greater than $\ellmax=300$ are utilized.
}
\label{fig:concal}}

The most important aspect of Fig.~\ref{fig:cal} is that the degradations are relatively small if 
we utilize all of the information to $\ellmax \gtrsim 10^3$.  Degradations are scale dependent, 
but consider the level of degradation at $\ellmax=3000$ for definiteness.  After calibrating 
halo structure, the $\wdist$ constraint is degraded by $\sim 34\% $ for LSST and roughly 
$\sim 18\% $ for DES and SNAP.  The $\wgrow$ constraint 
is degraded by $ \sim 18\% $ for LSST and less than $10\% $ for LSST and SNAP.  
The relevant combimation $\wdiff=\wdist-\wgrow$ is degraded by a maximum of 
$\sim 36\%$ for LSST and only about $\sim 13\%$ for SNAP and DES.  
To appreciate the utility of self-calibration in this instance, we should compare 
these numbers to the information loss associated with excising small scale information 
in Fig.~\ref{fig:wbiases}.  For LSST, excising high-$\ell$ information to eliminate biases 
requires taking $\ellmax \sim 300$ with a corresponding $\sim 300\% $ increase in the constraints 
on $\wdist$ and $\wgrow$.  So long as baryonic effects can be modeled in this way, the 
more sensible strategy is clearly to use all available information and calibrate 
the physics of galaxy formation as encoded in effective halo concentrations.  
The rightmost panel of Fig.~\ref{fig:cal} shows a similar result for the gravitational 
growth index $\gamma$.  In particular, $\sigma_{\gamma}$ is degraded by only $ \sim 20\% $ 
after self calibration as compared to $ \sim 200\% $ as would be required by the 
excision of small scales (see Fig.~\ref{fig:gbiases}).

For completeness, we give a revised set of constraint contours in the $\wdist$-$\wgrow$ 
plane in Fig.~\ref{fig:concal}.  After calibration, the contours are slightly expanded, 
but their orientation in this plane is only slightly changed.  By comparison, eliminating 
small-scale information corresponds to a dramatic loss of constraining power both because the 
area of the constraint contour in the $\wdist$-$\wgrow$ plane grows and because its orientation 
changes such that the most degenerate combination becomes more nearly perpindicular to the line 
$\wdist=\wgrow$.  The overall conclusion remains that self calibration is the appropriate strategy to adopt.

\section{Discussion}
\label{section:discussion}

We have extended recent studies of the influence of galaxy formation on weak lensing observables 
(in particular, Ref.~\cite{zentner_etal08}) 
to address cross-checks of the consistency of general relativity with weak lensing.  
Weak lensing is a key ingredient in such programs because weak lensing observables are 
sensitive to both the cosmological distance scale and the evolution of potential inhomogeneities, 
whereas the Type Ia supernovae distance-redshift relation and baryonic acoustic oscillation 
measurements are sensitive only to geometry (though it may be possible to exploit the 
lensing of superovae while performing the distance-redshift test, 
see Refs.~\cite{metcalf99,dodelson_vallinotto06,zentner_bhattacharya08} or to combine 
distance measures with cluster counts).  
Proposed methods to utilize weak lensing to constrain dark energy and to place 
limits on deviations from general relativity on cosmological scales assume that 
correlations on scales as small as a few arcminutes ($\ell \sim 10^3$) can be 
brought to bear on the cause of cosmic acceleration.  Such small-scale information 
is important to these programs.  Relegating consideration to larger scales can 
significantly degrade the constraining power of forthcoming experiments (see 
Fig.~\ref{fig:wbiases} and Fig.~\ref{fig:gbiases}) by  
nearly a factor of $\sim 3$ decrease in the effectiveness of tests for deviations from general relativity.

One of the anticipated challenges of using this small-scale information is that 
it is difficult to predict observables on these scales because the physics 
that governs the evolution of the baryonic component of the Universe is 
poorly understood and affects these observables in important ways 
\cite{white04,zhan_knox04,jing_etal06,rudd_etal08}.  
In Fig.~\ref{fig:wbiases} and Fig.~\ref{fig:gbiases}, we show that using such data to 
test general relativity may 
introduce potentially large biases in the inferred values of cosmological parameters 
that can lead to ruling out general relativity when it is in fact the true theory of gravitation.   
In these figures, we give a range of estimates of potential biases that may be realized.  These 
results are motivated by the specific results of the simulations of \cite{rudd_etal08}.  
The biases can be many times the statistical uncertainties in these parameters, which 
is clearly a serious problem.  We reiterate here that we give a range of biases that may 
be realized precisely because it is difficult to predict the precise influences of 
baryons.  We have used simulation results as guidance and explored a range of 
concentration relations that are consistent with current observations 
(e.g., Ref.~\cite{mandelbaum_etal08}).  We have not included information 
from additional observables that can be brought to bear on this problem.  For example, 
supernova Ia and baryon acoustic oscillation data can constrain the cosmic distance-redshift 
relation and be combined with lensing to produce 
stronger consistency checks.  However, these additional constraints on cosmological 
distances would only make the discord between structure growth and distance measures 
more egregious and drive larger biases, making the problem more severe.

Faced with a large systematic error, a common practice is to degrade the experiment in some way 
so as to derive robust results from the data and reduce sensitivity to the systematic error.  
Baryonic influences are scale dependent, so one way to degrade the experiment and 
eliminate these biases is to disregard small-scale information by restricting the analysis to 
$\ellmax \sim \mathrm{a}\ \mathrm{few}\times 10^2$.  
However, in Fig.~\ref{fig:wbiases} we have shown that this comes at the cost of nearly a factor of 
$\sim 3$ decrease in the effectiveness of tests for deviations from general relativity (more 
sophisticated analyses, such as those in Ref.~\cite{huterer_white05}, yield similar results).

As an alternative to excising small scale information, simulations suggest that we can understand 
the form of the baryonic influence and correct for it using a parameterized model.  To the 
degree that this form is not exactly degenerate with parameterizations of dark energy or 
deviations from general relativity, this enables the utilization of some of the information 
on small scales.  We addressed these biases by including $3$ baryonic parameters, 
$\alpha, \beta,$ and $c_0,$ that model the effective concentration of dark matter halos as 
specified in Eq.~\ref{eq:cofm}.  This choice of parametrization is motivated by 
Refs.~\cite{rudd_etal08,zentner_etal08}, wherein the authors demonstrated 
that modifying halo concentrations allows one to model account for the net 
influences of baryonic processes on scales relevant to weak lensing parameter 
estimation.  In particular, models in which the $N$-body results are corrected 
by a modified halo concentration relation track the convergence spectra from 
galaxy formation simulations out to $\ell \sim 5000$ accurately enough to reduce the biases 
in inferred dark energy parameters to levels below the statistical uncertainty of 
forthcoming experiments.  Within this framework, we studied the ability of 
forthcoming imaging surveys to self-calibrate the values of the baryonic parameters 
and simultaneously use this information to check the consistency of general relativity.

In Fig.~\ref{fig:cal} we have illustrated the effectiveness of our method by using the 
Fisher Matrix formalism to estimate the constraints on $\gamma$ and $w_g-w_d$ that will 
be obtained from future photometric surveys.   From this figure, it is evident that the 
constraints on the consistency of general relativity degrade by just $\sim 30\%$ relative 
to the limit of perfect knowledge of nonlinear evolution of the gravitational potential.  
In Fig.~\ref{fig:concal} the constraints on the consistency of general relativity obtained 
through our methods are much tighter than those from analyses excising small scale information.  
In light of our findings, we conclude that self-calibration of some of the specific influences 
of baryonic physics may be an appealing alternative to disregarding the information 
contained on scales smaller than $\ell \sim 300$.

Our outlook for self-calibration of uncertain baryonic physics is quite positive; however, there 
are several important caveats to our study and a number of additional studies that must 
be undertaken in order to bring such a program to fruition.  First, this correction for the physics of galaxy 
formation has only been applied to the simulations of Ref.~\cite{rudd_etal08}.  As the 
physics of baryons is highly uncertain, a more comprehensive exploration of viable alternative 
models is needed in order to validate such a correction term.  In fact, it is not reasonable 
to suppose that the influences of baryons should be strictly confined to the internal 
structures of halos (see Ref.~\cite{stanek_etal08} for changes in the halo mass function) 
and other effects may prove important.  Nevertheless, our study indicates that such 
an exploration is a fruitful pursuit.

Second, any such tests have important limitations.  The self-calibration program that 
we have explored should guard against the possibility of ruling out general relativity 
when it is the correct theory of gravity.  In our particular study, we have not 
explored complete, self-consistent phenomenologies for modified gravity.  Rather, 
we have only explored proposals that are already in the literature in which 
additional parameters are introduced which have known, fixed values in 
general relativity and explored the ability of forthcoming surveys to constrain 
these parameters.  In the event of a detection of a deviation from general relativity 
such tests provide limited information about the character of the correct, alternative theory.  
In large part, this is due to the fact that modified gravity models must have some 
scale dependence so that they can simultaneously drive accelerated cosmic expansion 
yet satisfy small-scale bounds on the theory of gravity~\cite{khoury_weltman04a,khoury_weltman04b,
navarro_acoleyen06,navarro_acoleyen07,faulkner_etal07,hu_sawicki07a,oyaizu_etal08}.  If a modified gravity 
theory, including nonlinear evolution, were fully specified and explored 
and if galaxy formation processes were more completely understood, it would 
be possible to draw more specific conclusions from forthcoming imaging survey data.

Lastly, we note that we operate under the assumption that 
dissipationless $N$-body simulations will effectively calibrate 
the properties of the lensing field absent baryonic effects.  
In our halo model approach, this is tantamount to a calibration of dark matter 
halo abundance and halo clustering that is so accurate as to introduce negligible 
uncertainties.  This is a common premise to all such studies of dark energy 
constraints as it is known that the density field is not yet calibrated with 
sufficient accuracy to analyze forthcoming data 
\cite{huterer_takada05,tinker_etal08,robertson_etal08}.  
Our outlook is that it is plausible that a brute force 
simulation campaign can achieve sufficient accuracy in this regard, though it is certainly possible
that reaching this goal may be complicated by several factors that have not yet been thoroughly studied.
For example, the phenomenologies we explore here are relatively simple, as they treat modified gravity only as a 
modification to the growth function.  Ultimately, it will be of interest to 
study more specific and more complete phenomenological models of modified gravity 
that may lead to large corrections to the density field on scales of several Mpc.  
Additionally, even in the absence of corrections necessitated by modified gravity it is
 possible that a program of $N$-body simulations
 coupled with a calibration of baryonic effects 
may encounter unforeseen obstacles.  
In either case, the assumption that an adequate suite of $N$-body simulations will 
address the density field (in our parameterization halo bias and halo abundance) with 
sufficient precision so as to render errors in the predicted 
density field negligible in the dissipationless case may not be valid.  
In such cases, it would be necessary to allow for adequate 
model freedom to account for this uncertainty.  
In the language of our paper, this might amount to internal calibration of halo 
abundance and halo bias (presumably with significant prior information) as well.  
Such additional freedom would lead to further degradation in dark energy and/or 
modified gravity parameters.  We have chosen to limit the scope of 
our work along these lines and not studied this more complicated case.

\section{Summary}
\label{section:conclusions}

We have explored the ability of forthcoming weak lensing surveys to perform a 
consistency check on general relativity as the correct theory of gravity in 
light of recent uncertainties in predicting lensing observables on scales 
smaller than $\sim 10$~arcmin due to the poorly-understood evolution of 
the baryonic component of the Universe.  
After a pedagogical demonstration of the utility of such 
consistency checks in \S~\ref{sub:consistency}, we present the 
following results.

\begin{itemize}

\item[1.] Conducting a study of the consistency of general relativity by analyzing weak lensing 
observables from a forthcoming imaging survey (such as DES, JDEM/SNAP, or LSST) out to scales as 
small as $\ell \sim 10^3$ could potentially lead to inferred cosmological parameters that 
are biased by many times the statistical uncertainty of such surveys if baryonic effects 
on the nonlinear evolution of the gravitational potential are ignored.  The problem is 
severe in that {\em they could lead to rejection of the null hypothesis of general 
relativistic gravity, even when it is the true theory}.  Specific results 
depend upon the assumed survey properties and the importance of baryonic processes.

\item[2.]  Disregarding small-scale information beyond 
$\ell \sim \mathrm{a}\ \mathrm{few} \times 10^2$ 
significantly reduces these biases because of the 
strong scale-dependence of the influence of baryonic 
physics on weak lensing observables.  However, excising 
this information degrades the constraining power of forthcoming 
imaging surveys by a factor of $\sim 3$.

\item[3.] As an alternative to excising small-scale information, we explore a 
program of self-calibrating a phenomenological model of baryonic processes that 
is motivated by recent numerical simulations.  With such a model small-scale information out 
to  $\ell \sim 10^3$ can be exploited, and biases can be reduced to levels below those of 
the statistical uncertainty of forthcoming experiments.  
As shown in Fig.~\ref{fig:cal} and Fig.~\ref{fig:concal}, 
this method provides substantially tighter constraints on the consistency of general 
relativity than analyses that disregard small-scale information.

\item[4.]  The program to calibrate the baryonic correction internally allows 
small-scale information to be used to constrain gravity and 
guards against ruling out general relativity when it is the true theory.  
In the case of internal calibration, the limits provided by general relativity 
consistency checks are degraded by only $\sim 30\%$ relative to the limits 
that would be achieved if nonlinear structure formation were perfectly understood.
Moreover, interesting limits on galaxy formation models are produced as a 
``byproduct.''

\end{itemize}

These results should be carefully considered in preparation for the 
weak lensing surveys that will be conducted in the next decade.  At the very least, 
it should be clear that issues regarding the baryonic influences on lensing observables 
will need to be dealt with, but that uncertainty in this regime may not necessarily 
have dramatic consequences on the ability of imaging surveys to illuminate the 
cause of cosmic acceleration.  It seems likely that a program of internal calibration of 
baryonic physics will allow lensing surveys to achieve stringent limits on gravity and 
dark energy.  In order to ensure that this is the case, a 
comprehensive theoretical program will need to be undertaken to better understand 
viable alternatives for the net influences of baryons on observable power spectra.



\acknowledgments

We thank Wayne Hu, Dragan Huterer, Manoj Kaplinghat, Arthur Kosowsky, 
Andrey Kravtsov, Jeff Newman, Douglas Rudd, and Mel Sanger 
for useful discussions.  ARZ and APH are supported by 
the University of Pittsburgh, by the US National Science Foundation 
through grant AST 0806367, and by the US Department of Energy.


\bibliography{wcons}


\end{document}